\documentclass[prd,nofootinbib,preprint, 10 pt]{revtex4}
\usepackage{amssymb}
\usepackage{amsmath,graphicx,color,epsfig}
\usepackage{pstricks}
\usepackage{float}

\begin{document}

\title{ Semileptonic charmed $B$ meson decays in Universal Extra Dimension
Model}
\author{M. Ali Paracha}
\email{paracha@phys.qau.edu.pk}
\affiliation{Physics Department, Quaid-i-Azam University, Islamabad, Pakistan}
\author{Ishtiaq Ahmed}
\email{ishtiaq@ncp.edu.pk}
\affiliation{National Centre for Physics, Quaid-i-Azam University, Islambad, Pakistan}
\author{M. Jamil Aslam}
\email{jamil@phys.qau.edu.pk}
\affiliation{Physics Department, Quaid-i-Azam University, Islamabad, Pakistan.}
\date{\today }

\begin{abstract}
Form factors parameterizing the semileptonic decay $B_{c}\rightarrow
D_{s}^{\ast }l^{+}l^{-}$ ($l=\mu ,\tau $) are calculated using the frame
work of Ward Identities. These form factors are then used to calculate the
physical observables like branching ratio and helicity fractions of final
state $D_{s}^{\ast}$ meson in these decay modes. The analysis is then
extended to the the universal extra dimension (UED) model where the
dependence of above mentioned physical variables to the compactification
radius R, the only unknown parameter in UED model, is studied. It is shown
that the helicity fractions of $D_{s}^{\ast}$ are quite sensitive to the UED
model especially when have muons as the final state lepton. Therefore, these
can serve as a useful tool to establish new physics predicted by the UED
model.
\end{abstract}

\maketitle


\section{\protect\bigskip Introduction}

Living in the LHC era, it is hoped to either verify the Standard Model (SM)
or to explore the properties of more accurate underlying theory that
describes the theory of weak scale. Flavor Changing Neutral Current (FCNC)
decays of $B$-meson are an important tool to investigate the structure of
weak interactions and also provide us a frame work to look for the physics
beyond the Standard Model (SM). This lies in the fact that FCNC decays are
not allowed at tree level in the SM and occur only at the loop level \cite%
{1,2,2a} and makes them quite sensitive to possible small corrections that
may be result of any modification to the SM, or from the new interactions.
This gives us solid reason to study these decays both theoretically and
experimentally.

Since the CLEO observations of the rare radiative $b\rightarrow s\gamma $
transition \cite{3}, there have been intensive studies on rare semileptonic,
radiative and leptonic decays of $B_{u,d,s}$ mesons induced by FCNC
transitions of $b\rightarrow s,d$ \ \cite{4}. The study will be even more
complete if one consider the similar decays of the charmed $B$ mesons $%
(B_{c})$.

The charmed $B_{c}$ meson is a bound state of two heavy quarks, bottom $b$
and charm $c$, and was first observed in 1998 at Tevatron in Fermilab \cite%
{8}. Because of two heavy quarks, the $B_{c}$ mesons are rich in
phenomenology compared to the other $B$ mesons. At the Large Hadron Collider
(LHC) the expected number of events for the production of $B_{c}$ \ meson
are about $10^{8}-10^{10}$ per year \cite{9,10} which is a reasonable number
to work on the phenomenology of the $B_{c}$ meson. In literature, some of
the possible radiative and semileptonic exclusive decays of $B_{c}$ mesons
like $B_{c}\rightarrow \left( \rho ,K^{\ast },D_{s}^{\ast },B_{u}^{\ast
}\right) \gamma ,B_{c}\rightarrow l\nu \gamma $ $,B_{c}\rightarrow
B_{u}^{\ast }l^{+}l^{-},B_{c}\rightarrow D_{1}^{0}l\nu ,B_{c}\rightarrow
D_{s0}^{\ast }l^{+}l^{-}$ and $B_{c}\rightarrow D_{s,d}^{\ast }l^{+}l^{-}$
have been studied using the frame work of relativistic constituent quark
model \cite{11}, QCD Sum Rules and the Light Cone Sum Rules \cite{12}. The
focus of the present work is the study of exclusive $B_{c}\rightarrow
D_{s}^{\ast }l^{+}l^{-}$ decay.

While working on the exclusive $B$-meson decays the main job is to calculate
the form factors which are the non perturbative quantities and are the
scalar functions of the square of momentum transfer. In literature the form
factors for $B_{c}\rightarrow D_{s}^{\ast }l^{+}l^{-}$ decay were calculated
using different approaches, such as light front constituent quark models and
a relativistic quark model \cite{11, 13}. In this work we calculate the form
factors for the above mentioned decay in a model independent way through
Ward identities, which was earlier applied to $B\rightarrow \rho,\gamma $
\cite{14, MJAR} and $B\rightarrow K_{1}$ decays \cite{15}. This approach
enables us to make a clear separation between the pole and non pole type
contributions, the former is known in terms of a universal function $\xi
_{\perp }(q^{2})\equiv g_{+}(q^{2})$ which is introduced in the Large Energy
Effective Theory (LEET) of heavy to light transition form factors \cite{16}.
The residue of the pole is then determined in a self consistent way in terms
of $g_{+}(0)$ which will give information about the couplings of $%
B_{s}^{\ast }(1^{-})$ and $B_{sA}^{\ast }(1^{+})$ with $B_{c}D_{s}^{\ast }$
channel. The above mentioned coupling arises at lower pole masses because
the higher pole masses of $B_{c}$ meson do not contribute for the decay $%
B_{c}\rightarrow D_{s}^{\ast }l^{+}l^{-}.$ The form factors are then
determine in terms of a known parameter $g_{+}(0)$ and the pole masses of
the particles involved, which will then be used to calculate different
physical observables like the branching ratio and the helicity fractions of $%
D_{s}^{\ast }$ for these decays.

At the quark level the semileptonic decay $B_{c}\rightarrow D_{s}^{\ast
}l^{+}l^{-}$ is governed by the FCNC transition $b\rightarrow sl^{+}l^{-},$
therefore it is an important candidate to look for physics in and beyond the
SM. Many investigations for the physics beyond the SM are now being
performed in various areas of particle physics which are expected to get the
direct or indirect evidence at high energy colliders such as LHC. During the
last couple of years there have been an increased interest in models with
extra dimensions, since they solve the hierarchy problem and they can
provide the unified framework of gravity and other interactions together
with a connection to the string theory \cite{17}. Among them the special
role plays the one with universal extra dimensions (UED) as in this model
all SM\ fields are allowed to propagate in available all dimensions. The
economy of UED\ model is that there is only one additional parameter to that
of SM which is the radius $R$ of the compactified extra dimension. Now above
the compactification scale $1/R$ a given UED model becomes a higher
dimensional field theory whose equivalent description in four dimensions
consists of SM\ fields and the towers of KK modes having no partner in the
SM. A simplest model of this type was proposed by Appelquist,Cheng and
Dobrescu (ACD) \cite{18}. In this model all the masses of the KK particles
and their interactions with SM\ particles and also among themselves are
described in terms of the inverse of compactification radius $R$ and the
parameters of the SM \cite{19}.

The most important property of ACD model is the conservation of parity which
implies the absence of tree level contributions of KK states to the low
energy processes taking place at scale $\mu <<1/R.$ This brings interest
towards the FCNC\ transitions, $b\rightarrow s$ as mentioned earlier that
these transitions occur at loop level in SM and hence the one loop
contribution due to KK modes to them could in principly be important. These
processes are used to constrain the mass and couplings of the KK states,
i.e, the compactification radius $1/R$ \cite{19, 20}.\

Buras \textit{et al.} have computed the effective Hamiltonian of several
FCNC processes in ACD model, particularly in $b$ sector, namely $B_{s,d}$
mixing and $b\rightarrow s$ transition such as $b\rightarrow s\gamma $ and $%
b\rightarrow sl^{+}l^{-}$ decay \cite{19}. The implications of physics with
UED are examined with data from Tevatron experiments and the bounds on the
inverse of compactification radius are found to be $1/R\geq 250-300$ GeV
\cite{21}. There exists some studies in the literature on different $B$ to
light meson decays in ACD model, where the dependence of different physical
observables like branching ratio, forward-backward asymmetry, lepton
polarization asymmetry and the helicity fractions of final state mesons on $%
1/R$ is examined \cite{21, 22, 23}.

In this work we will study the branching ratio and helicity fractions of $%
D_{s}^{\ast }$ meson in $B_{c}\rightarrow D_{s}^{\ast }l^{+}l^{-}$ decay
both in the SM and ACD model using the framework of $B\rightarrow (K^{\ast
},K_{1})l^{+}l^{-}$ decays described in refs. \cite{22,23}. \ The paper is
organized as follows. In Sec. II we present the effective Hamiltonian for
the decay $B_{c}\rightarrow D_{s}^{\ast }l^{+}l^{-}.$ Section III contains
the definitions as well as the detailed calculation of the form factors using
Ward Identities. In Sec. IV we present the basic formulas for physical
observables like decay rate and helicity fractions of $D_{s}^{\ast }$ meson
where as the numerical analysis of these observables will be given in
Section V. Section VI gives the summary of the results.

\section{Effective Hamiltonian and Matrix Elements}

At quark level, the semileptonic decay $B_{c}\rightarrow D_{s}^{\ast
}l^{+}l^{-}$ is governed by the transition $b\rightarrow sl^{+}l^{-}$ for
which the effective Hamiltonian can be written as
\begin{eqnarray}
H_{eff}&=&-\frac{4G_{F}}{\sqrt{2}}V_{tb}V_{ts}^{\ast }\bigg[ %
\sum\limits_{i=1}^{10}C_{i}(\mu )O_{i}\bigg],  \label{1}
\end{eqnarray}%
where $O_{i}(\mu )$ $(i=1,...,10)$ are the four quark operators and $%
C_{i}(\mu )$ are the corresponding Wilson coefficients at the energy scale $%
\mu $ \cite{24} which was usually take to be the $b$-quark mass $\left(
m_{b}\right) $. The theoretical uncertainties related to the
renormalization scale can be reduced when the next to leading logarithm
corrections are included. The explicit form of the operators responsible for
the decay $B_{c}^{-}\rightarrow D_{s}^{\ast -}l^{+}l^{-}$ is%
\begin{eqnarray}
O_{7} &=&\frac{e^{2}}{16\pi ^{2}}m_{b}(\bar{s}\sigma _{\mu \nu }Rb)F^{\mu
\nu }  \label{2} \\
O_{9} &=&\frac{e^{2}}{16\pi ^{2}}\left( \bar{s}\gamma _{\mu }Lb\right) \bar{l%
}\gamma ^{\mu }l  \label{3} \\
O_{10} &=&\frac{e^{2}}{16\pi ^{2}}\left( \bar{s}\gamma _{\mu }Lb\right) \bar{%
l}\gamma ^{\mu }\gamma ^{5}l  \label{4}
\end{eqnarray}%
with $L,R=\left( 1\mp \gamma ^{5}\right) /2$.

Using the effective Hamiltonian given in Eq.(\ref{1}) the free quark
amplitude for $b\rightarrow sl^{+}l^{-}$ can be written as%
\begin{eqnarray}
\mathcal{M(}b\rightarrow sl^{+}l^{-})&=&-\frac{G_{F}\alpha }{\sqrt{2}\pi }%
V_{tb}V_{ts}^{\ast }\bigg[{C_{9}^{eff}\left( \mu \right) (\bar{s}\gamma
_{\mu }Lb)(\bar{l}\gamma ^{\mu}l)+C_{10}(}\bar{s}\gamma _{\mu }Lb)(\bar{l}%
\gamma ^{\mu }\gamma ^{5}l)  \notag \\
&&-2C_{7}^{eff}\left( \mu \right) \frac{m_{b}}{q^{2}}(\bar{s}i\sigma _{\mu
\nu}q^{\nu }Rb)\bar{l}\gamma ^{\mu }l\bigg]  \label{5a}
\end{eqnarray}
where $q^{2}$ is the square of momentum transfer. Note that the operator $%
O_{10}$ \ given in Eq.(\ref{4}) can not be induced by the insertion of four
quark operators because of the absence of $Z$-boson in the effective theory.
Therefore, the Wilson coefficient $C_{10}$ does not renormalize under QCD
corrections and is independent on the energy scale $\mu .$ Additionally the
above quark level decay amplitude can get contributions from the matrix
element of four quark operators, $\sum_{i=1}^{6}\left\langle
l^{+}l^{-}s\left\vert O_{i}\right\vert b\right\rangle ,$ which are usually
absorbed into the effective Wilson coefficient $C_{9}^{eff}(\mu )$ and can
be written as \cite{25, 26, 27, 28, 29, 30, 31}%
\begin{equation*}
C_{9}^{eff}(\mu )=C_{9}(\mu )+Y_{SD}(z,s^{\prime })+Y_{LD}(z,s^{\prime }).
\end{equation*}%
where $z=m_{c}/m_{b}$ and $s^{\prime }=q^{2}/m_{b}^{2}$. $Y_{SD}(z,s^{\prime
})$ describes the short distance contributions from four-quark operators far
away from the $c\bar{c}$ resonance regions, and this can be calculated
reliably in the perturbative theory. However the long distance contribution $%
Y_{LD}(z,s^{\prime })$ cannot be calculated by using the first principles of
QCD, so they are usually parameterized in the form of a phenomenological
Breit-Wigner formula making use of the vacuum saturation approximation and
quark hadron duality. Therefore, one can not calculate them reliably so we
we will neglect these long distance effects for the case of $%
B_{c}\rightarrow D_{s}^{\ast }l^{+}l^{-}$.  The expression for the short
distance contribution $Y_{SD}(z,s^{\prime })$ is given as
\begin{eqnarray}
Y_{SD}(z,s^{\prime }) &=&h(z,s^{\prime })(3C_{1}(\mu )+C_{2}(\mu
)+3C_{3}(\mu )+C_{4}(\mu )+3C_{5}(\mu )+C_{6}(\mu ))  \notag \\
&&-\frac{1}{2}h(1,s^{\prime })(4C_{3}(\mu )+4C_{4}(\mu )+3C_{5}(\mu
)+C_{6}(\mu ))  \notag \\
&&-\frac{1}{2}h(0,s^{\prime })(C_{3}(\mu )+3C_{4}(\mu ))+{\frac{2}{9}}%
(3C_{3}(\mu )+C_{4}(\mu )+3C_{5}(\mu )+C_{6}(\mu )),
\end{eqnarray}
with
\begin{eqnarray}
h(z,s^{\prime }) &=&-{\frac{8}{9}}\mathrm{ln}z+{\frac{8}{27}}+{\frac{4}{9}}x-%
{\frac{2}{9}}(2+x)|1-x|^{1/2}\left\{
\begin{array}{l}
\ln \left| \frac{\sqrt{1-x}+1}{\sqrt{1-x}-1}\right| -i\pi \quad \mathrm{for}{%
{\ }x\equiv 4z^{2}/s^{\prime }<1} \\
2\arctan \frac{1}{\sqrt{x-1}}\qquad \mathrm{for}{{\ }x\equiv
4z^{2}/s^{\prime }>1}%
\end{array}
\right. ,  \notag \\
h(0,s^{\prime }) &=&{\frac{8}{27}}-{\frac{8}{9}}\mathrm{ln}{\frac{m_{b}}{\mu
}}-{\frac{4}{9}}\mathrm{ln}s^{\prime }+{\frac{4}{9}}i\pi \,\,.
\end{eqnarray}
Also the non factorizable effects from the charm loop brings further
corrections to the radiative transition $b\rightarrow s\gamma ,$ and these
can be absorbed into the effective Wilson coefficients $C_{7}^{eff}$ which
then takes the form \cite{32, 33, 34, 35, 36}
\begin{equation*}
C_{7}^{eff}(\mu )=C_{7}(\mu )+C_{b\rightarrow s\gamma }(\mu )
\end{equation*}%
with
\begin{eqnarray}
C_{b\rightarrow s\gamma }(\mu ) &=&i\alpha _{s}\left[ \frac{2}{9}\eta
^{14/23}(G_{1}(x_{t})-0.1687)-0.03C_{2}(\mu )\right]  \label{8} \\
G_{1}(x_{t}) &=&\frac{x_{t}\left( x_{t}^{2}-5x_{t}-2\right) }{8\left(
x_{t}-1\right) ^{3}}+\frac{3x_{t}^{2}\ln ^{2}x_{t}}{4\left( x_{t}-1\right)
^{4}}  \label{9}
\end{eqnarray}%
where $\eta =\alpha _{s}(m_{W})/\alpha _{s}(\mu ),$ \ $%
x_{t}=m_{t}^{2}/m_{W}^{2}$ and $C_{b\rightarrow s\gamma }$ is the absorptive
part for the $b\rightarrow sc\bar{c}\rightarrow s\gamma $ rescattering.

The new physics effects manifest themselves in rare $B$ decays in two
different ways, either through new contribution to the Wilson coefficients
or through the new operators in the effective Hamiltonian, which are absent
in the SM. Being minimal extension of SM the ACD model is the most
economical one because it has only additional parameter $R$ i.e. the radius
of the compactification leaving the operators basis same as that of the SM.
Therefore, the whole contribution from all the KK states is in the Wilson
coefficients which are now the functions of the compactification radius $R$.
At large value of $1/R$ the new states being more and more massive and will
be decoupled from the low-energy theory,therefore one can recover the SM
phenomenology.

The modified Wilson coefficients in ACD model contain the contribution from
new particles which are not present in the SM and comes as an intermediate
state in penguin and box diagrams. Thus, these coefficients can be expressed
in terms of the functions $F\left( x_{t},1/R\right) $, $x_{t}=\frac{m_{t}^{2}%
}{M_{W}^{2}}$, which generalize the corresponding SM\ function $F_{0}\left(
x_{t}\right) $ according to:
\begin{equation}
F\left( x_{t},1/R\right) =F_{0}\left( x_{t}\right) +\sum_{n=1}^{\infty
}F_{n}\left( x_{t},x_{n}\right)  \label{f-expression}
\end{equation}%
with $x_{n}=\frac{m_{n}^{2}}{M_{W}^{2}}$ and $m_{n}=\frac{n}{R}$ \cite{44}.
The relevant diagrams are $Z^{0}$ penguins, $\gamma $ penguins, gluon
penguins, $\gamma $ magnetic penguins, Chormomagnetic penguins$\ $and the
corresponding functions are $C\left( x_{t},1/R\right) $, $D\left(
x_{t},1/R\right) $, $E\left( x_{t},1/R\right) $, $D^{\prime }\left(
x_{t},1/R\right) $ and $E^{\prime }\left( x_{t},1/R\right) $ respectively.
These functions are calculated at next to leading order by Buras et al. \cite%
{19} and can be summarized as:

$\bullet C_{7}$

In place of $C_{7},$ one defines an effective coefficient $C_{7}^{(0)eff}$
which is renormalization scheme independent \cite{45}:
\begin{equation}
C_{7}^{(0)eff}(\mu _{b})=\eta ^{\frac{16}{23}}C_{7}^{(0)}(\mu _{W})+\frac{8}{%
3}(\eta ^{\frac{14}{23}}-\eta ^{\frac{16}{23}})C_{8}^{(0)}(\mu
_{W})+C_{2}^{(0)}(\mu _{W})\sum_{i=1}^{8}h_{i}\eta ^{\alpha _{i}}
\label{wilson1}
\end{equation}%
where $\eta =\frac{\alpha _{s}(\mu _{W})}{\alpha _{s}(\mu _{b})}$, and
\begin{equation}
C_{2}^{(0)}(\mu _{W})=1,\mbox{ }C_{7}^{(0)}(\mu _{W})=-\frac{1}{2}D^{\prime
}(x_{t},\frac{1}{R}),\mbox{
}C_{8}^{(0)}(\mu _{W})=-\frac{1}{2}E^{\prime }(x_{t},\frac{1}{R});
\label{wilson2}
\end{equation}%
the superscript $(0)$ stays for leading logarithm approximation.
Furthermore:
\begin{eqnarray}
\alpha _{1} &=&\frac{14}{23}\mbox{ \quad }\alpha _{2}=\frac{16}{23}%
\mbox{
\quad }\alpha _{3}=\frac{6}{23}\mbox{ \quad
}\alpha _{4}=-\frac{12}{23}  \notag \\
\alpha _{5} &=&0.4086\mbox{ \quad }\alpha _{6}=-0.4230\mbox{ \quad
}\alpha _{7}=-0.8994\mbox{ \quad }\alpha _{8}=-0.1456  \notag \\
h_{1} &=&2.996\mbox{ \quad }h_{2}=-1.0880\mbox{ \quad }h_{3}=-\frac{3}{7}%
\mbox{ \quad }h_{4}=-\frac{1}{14}  \notag \\
h_{5} &=&-0.649\mbox{ \quad }h_{6}=-0.0380\mbox{ \quad
}h_{7}=-0.0185\mbox{ \quad }h_{8}=-0.0057.  \label{wilson3}
\end{eqnarray}%
The functions $D^{\prime }$ and $E^{\prime }$ are
\begin{equation}
D_{0}^{\prime }(x_{t})=-\frac{(8x_{t}^{3}+5x_{t}^{2}-7x_{t})}{12(1-x_{t})^{3}%
}+\frac{x_{t}^{2}(2-3x_{t})}{2(1-x_{t})^{4}}\ln x_{t},  \label{wilson4}
\end{equation}
\begin{equation}
E_{0}^{\prime }(x_{t})=-\frac{x_{t}(x_{t}^{2}-5x_{t}-2)}{4(1-x_{t})^{3}}+%
\frac{3x_{t}^{2}}{2(1-x_{t})^{4}}\ln x_{t},  \label{wilson5}
\end{equation}
\begin{eqnarray}  \label{wilson6}
D_{n}^{\prime }(x_{t},x_{n}) &=&\frac{%
x_{t}(-37+44x_{t}+17x_{t}^{2}+6x_{n}^{2}(10-9x_{t}+3x_{t}^{2})-3x_{n}(21-54x_{t}+17x_{t}^{2}))%
}{36(x_{t}-1)^{3}}  \notag \\
&&+\frac{x_{n}(2-7x_{n}+3x_{n}^{2})}{6}\ln \frac{x_{n}}{1+x_{n}}  \notag \\
&&-\frac{%
(-2+x_{n}+3x_{t})(x_{t}+3x_{t}^{2}+x_{n}^{2}(3+x_{t})-x_{n})(1+(-10+x_{t})x_{t}))%
}{6(x_{t}-1)^{4}} \ln \frac{x_{n}+x_{t}}{1+x_{n}},  \notag \\
\end{eqnarray}
\begin{eqnarray}
E_{n}^{\prime }(x_{t},x_{n}) &=&\frac{%
x_{t}(-17-8x_{t}+x_{t}^{2}+3x_{n}(21-6x_{t}+x_{t}^{2})-6x_{n}^{2}(10-9x_{t}+3x_{t}^{2}))%
}{12(x_{t}-1)^{3}}  \notag \\
&&+-\frac{1}{2}x_{n}(1+x_{n})(-1+3x_{n})\ln \frac{x_{n}}{1+x_{n}}  \notag \\
&&+\frac{%
(1+x_{n})(x_{t}+3x_{t}^{2}+x_{n}^{2}(3+x_{t})-x_{n}(1+(-10+x_{t})x_{t}))}{%
2(x_{t}-1)^{4}}\ln \frac{x_{n}+x_{t}}{1+x_{n}}.  \label{wilson7}
\end{eqnarray}
Following reference \cite{19}, one gets the expressions for the sum over $n:$%
\begin{eqnarray}
\sum_{n=1}^{\infty }D_{n}^{\prime }(x_{t},x_{n}) &=&-\frac{%
x_{t}(-37+x_{t}(44+17x_{t}))}{72(x_{t}-1)^{3}}  \notag \\
&&+\frac{\pi M_{w}R}{2}[\int_{0}^{1}dy\frac{2y^{\frac{1}{2}}+7y^{\frac{3}{2}%
}+3y^{\frac{5}{2}}}{6}]\coth (\pi M_{w}R\sqrt{y})  \notag \\
&&+\frac{(-2+x_{t})x_{t}(1+3x_{t})}{6(x_{t}-1)^{4}}J(R,-\frac{1}{2})  \notag
\\
&&-\frac{1}{6(x_{t}-1)^{4}}%
[x_{t}(1+3x_{t})-(-2+3x_{t})(1+(-10+x_{t})x_{t})]J(R,\frac{1}{2})  \notag \\
&&+\frac{1}{6(x_{t}-1)^{4}}[(-2+3x_{t})(3+x_{t})-(1+(-10+x_{t})x_{t})]J(R,%
\frac{3}{2})  \notag \\
&&-\frac{(3+x_{t})}{6(x_{t}-1)^{4}}J(R,\frac{5}{2})],  \label{wilson8}
\end{eqnarray}
\begin{eqnarray}
\sum_{n=1}^{\infty }E_{n}^{\prime }(x_{t},x_{n}) &=&-\frac{%
x_{t}(-17+(-8+x_{t})x_{t})}{24(x_{t}-1)^{3}}  \notag \\
&&+\frac{\pi M_{w}R}{2}[\int_{0}^{1}dy(y^{\frac{1}{2}}+2y^{\frac{3}{2}}-3y^{%
\frac{5}{2}})\coth (\pi M_{w}R\sqrt{y})]  \notag \\
&&-\frac{x_{t}(1+3x_{t})}{(x_{t}-1)^{4}}J(R,-\frac{1}{2})  \notag \\
&&+\frac{1}{(x_{t}-1)^{4}}[x_{t}(1+3x_{t})-(1+(-10+x_{t})x_{t})]J(R,\frac{1}{%
2})  \notag \\
&&-\frac{1}{(x_{t}-1)^{4}}[(3+x_{t})-(1+(-10+x_{t})x_{t})]J(R,\frac{3}{2})
\notag \\
&&+\frac{(3+x_{t})}{(x_{t}-1)^{4}}J(R,\frac{5}{2})],  \label{wilson9}
\end{eqnarray}
where
\begin{equation}
J(R,\alpha )=\int_{0}^{1}dyy^{\alpha }[\coth (\pi M_{w}R\sqrt{y}%
)-x_{t}^{1+\alpha }\coth (\pi m_{t}R\sqrt{y})].  \label{wilson10}
\end{equation}
$\bullet C_{9}$

In the ACD model and in the NDR scheme one has
\begin{equation}
C_{9}(\mu )=P_{0}^{NDR}+\frac{Y(x_{t},\frac{1}{R})}{\sin ^{2}\theta _{W}}%
-4Z(x_{t},\frac{1}{R})+P_{E}E(x_{t},\frac{1}{R})  \label{wilson11}
\end{equation}%
where $P_{0}^{NDR}=2.60\pm 0.25$ \cite{12} and the last term is numerically
negligible. Besides
\begin{eqnarray}
Y(x_{t},\frac{1}{R}) &=&Y_{0}(x_{t})+\sum_{n=1}^{\infty }C_{n}(x_{t},x_{n})
\notag \\
Z(x_{t},\frac{1}{R}) &=&Z_{0}(x_{t})+\sum_{n=1}^{\infty }C_{n}(x_{t},x_{n})
\label{wilson12}
\end{eqnarray}%
with
\begin{eqnarray}
Y_{0}(x_{t}) &=&\frac{x_{t}}{8}[\frac{x_{t}-4}{x_{t}-1}+\frac{3x_{t}}{%
(x_{t}-1)^{2}}\ln x_{t}]  \notag \\
Z_{0}(x_{t}) &=&\frac{18x_{t}^{4}-163x_{t}^{3}+259x_{t}^{2}-108x_{t}}{%
144(x_{t}-1)^{3}}  \notag \\
&&+[\frac{32x_{t}^{4}-38x_{t}^{3}+15x_{t}^{2}-18x_{t}}{72(x_{t}-1)^{4}}-%
\frac{1}{9}]\ln x_{t}  \label{wilson13}
\end{eqnarray}%
\begin{equation}
C_{n}(x_{t},x_{n})=\frac{x_{t}}{8(x_{t}-1)^{2}}%
[x_{t}^{2}-8x_{t}+7+(3+3x_{t}+7x_{n}-x_{t}x_{n})\ln \frac{x_{t}+x_{n}}{%
1+x_{n}}]  \label{wilson14}
\end{equation}%
and
\begin{equation}
\sum_{n=1}^{\infty }C_{n}(x_{t},x_{n})=\frac{x_{t}(7-x_{t})}{16(x_{t}-1)}-%
\frac{\pi M_{w}Rx_{t}}{16(x_{t}-1)^{2}}[3(1+x_{t})J(R,-\frac{1}{2}%
)+(x_{t}-7)J(R,\frac{1}{2})]  \label{wilson15}
\end{equation}%
$\bullet C_{10}$

$C_{10}$ is $\mu $ independent and is given by
\begin{equation}
C_{10}=-\frac{Y(x_{t},\frac{1}{R})}{\sin ^{2}\theta _{w}}.  \label{wilson16}
\end{equation}%
The normalization scale is fixed to $\mu =\mu _{b}\simeq 5$ GeV.

\section{Matrix Elements and Form Factors}

The exclusive $B_{c}\rightarrow D_{s}^{\ast }l^{+}l^{-}$decay involves the
hadronic matrix elements which can be obtained by sandwiching the quark
level operators give in Eq. (\ref{5a}) between initial state $B_{c}$ meson
and final state $D_{s}^{\ast }$ meson. These can be parameterized in terms of
form factors which are scalar functions of the square of the four momentum
transfer($q^{2}=(p-k)^{2}).$ The non vanishing matrix elements for the
process $B_{c}\rightarrow D_{s}^{\ast }$ can be parameterized in terms of
the seven form factors as follows%
\begin{eqnarray}
\left\langle D_{s}^{\ast }(k,\varepsilon )\left\vert \bar{s}\gamma _{\mu
}b\right\vert B_{c}(p)\right\rangle &=&\frac{2\epsilon _{\mu \nu \alpha
\beta }}{M_{B_{c}}+M_{D_{s}^{\ast -}}}\varepsilon ^{\ast \nu }p^{\alpha
}k^{\beta }V(q^{2})  \label{10} \\
\left\langle D_{s}^{\ast }(k,\varepsilon )\left\vert \bar{s}\gamma _{\mu
}\gamma _{5}b\right\vert B_{c}(p)\right\rangle &=&i\left(
M_{B_{c}^{-}}+M_{D_{s}^{\ast -}}\right) \varepsilon ^{\ast \mu }A_{1}(q^{2})
\notag \\
&&-i\frac{(\varepsilon ^{\ast }\cdot q)}{M_{B_{c}^{-}}+M_{D_{s}^{\ast -}}}%
\left( p+k\right) ^{\mu }A_{2}(q^{2})  \notag \\
&&-i\frac{2M_{D_{s}^{\ast -}}}{q^{2}}q^{\mu }(\varepsilon ^{\ast }\cdot q)%
\left[ A_{3}(q^{2})-A_{0}(q^{2})\right]  \notag \\
&&  \label{11}
\end{eqnarray}%
where $p$ is the momentum of $B_{c}$, $\varepsilon $ and $k$ are the
polarization vector and momentum of the final state $D_{s}^{\ast }$ meson.
Here, the form factor $A_{3}(q^{2})$ can be expressed in terms of the form
factors $A_{1}(q^{2})$ and $A_{2}(q^{2})$ as
\begin{equation}
A_{3}(q^{2})=\frac{M_{B_{c}^{-}}+M_{D_{s}^{\ast -}}}{2M_{D_{s}^{\ast -}}}%
A_{1}(q^{2})-\frac{M_{B_{c}^{-}}-M_{D_{s}^{\ast -}}}{2M_{D_{s}^{\ast -}}}%
A_{2}(q^{2})  \label{12a}
\end{equation}%
with
\begin{equation*}
A_{3}(0)=A_{0}(0)
\end{equation*}%
In addition to the above form factors there are some penguin form factors,
which we can write as
\begin{eqnarray}
\left\langle D_{s}^{\ast }(k,\varepsilon )\left\vert \bar{s}i\sigma _{\mu
\nu }q^{\nu }b\right\vert B_{c}(p)\right\rangle &=&-\epsilon _{\mu \nu
\alpha \beta }\varepsilon ^{\ast \nu }p^{\alpha }k^{\beta }2F_{1}(q^{2})
\label{13a} \\
\left\langle D_{s}^{\ast }(k,\varepsilon )\left\vert \bar{s}i\sigma _{\mu
\nu }q^{\nu }\gamma ^{5}b\right\vert B_{c}(p)\right\rangle &=&i\left[ \left(
M_{Bc^{-}}^{2}-M_{D_{s}^{\ast -}}^{2}\right) \varepsilon _{\mu
}-(\varepsilon ^{\ast }\cdot q)(p+k)_{\mu }\right] F_{2}(q^{2})  \notag \\
&&  \label{13b} \\
&&+(\varepsilon ^{\ast }\cdot q)i\left[ q_{\mu }-\frac{q^{2}}{%
M_{Bc^{-}}^{2}-M_{D_{s}^{\ast -}}^{2}}(p+k)_{\mu }\right] F_{3}(q^{2})
\notag
\end{eqnarray}%
with
\begin{equation*}
F_{1}(0)=F_{2}(0)
\end{equation*}

Now the different form factors appearing in Eqs. (\ref{10}-\ref{13b}) can be
related to each other with the help of Ward identities as follows \cite{14}%
\begin{eqnarray}
\left\langle D_{s}^{\ast }(k,\varepsilon )\left\vert \bar{s}i\sigma _{\mu
\nu }q^{\nu }b\right\vert B_{c}(p)\right\rangle &=&(m_{b}+m_{s})\left\langle
D_{s}^{\ast }(k,\varepsilon )\left\vert \bar{s}\gamma _{\mu }b\right\vert
B_{c}(p)\right\rangle  \label{14a} \\
\left\langle D_{s}^{\ast }(k,\varepsilon )\left\vert \bar{s}i\sigma _{\mu
\nu }q^{\nu }\gamma ^{5}b\right\vert B_{c}(p)\right\rangle
&=&-(m_{b}-m_{s})\left\langle D_{s}^{\ast }(k,\varepsilon )\left\vert \bar{s}%
\gamma _{\mu }\gamma _{5}b\right\vert B_{c}(p)\right\rangle  \notag \\
&&+(p+k)_{\mu }\left\langle D_{s}^{\ast }(k,\varepsilon )\left\vert \bar{s}%
\gamma _{5}b\right\vert B_{c}(p)\right\rangle  \label{14b}
\end{eqnarray}%
By putting Eq.(\ref{10}-\ref{13b}) in Eq.(\ref{14a}) and (\ref{14b}) and
comparing the coefficients of $\varepsilon _{\mu }^{\ast }$ and $q_{\mu }$
on both sides, one can get the following relations between the form factors:%
\begin{eqnarray}
F_{1}(q^{2}) &=&\frac{(m_{b}+m_{s})}{M_{B_{c}^{-}}+M_{D_{s}^{\ast -}}}%
V(q^{2})  \label{13} \\
F_{2}(q^{2}) &=&\frac{m_{b}-m_{s}}{M_{B_{c}^{-}}+M_{D_{s}^{\ast -}}}%
A_{1}(q^{2})  \label{14} \\
F_{3}(q^{2}) &=&-(m_{b}-m_{s})\frac{2M_{D_{s}^{\ast -}}}{q^{2}}\left[
A_{3}(q^{2})-A_{0}(q^{2})\right]  \label{15}
\end{eqnarray}%
The results given in Eqs. (\ref{13}, \ref{14}, \ref{15}) are derived by
using Ward identities and therefore are the model independent.

The universal normalization of the above form factors at $q^{2}=0$ are
obtained by defining \cite{14}%
\begin{eqnarray}
\left\langle D_{s}^{\ast }(k,\varepsilon )\left\vert \bar{s}i\sigma _{\alpha
\beta }b\right\vert B_{c}(p)\right\rangle &=&-i\epsilon _{\alpha \beta \rho
\sigma }\varepsilon ^{\ast \rho }\left[ (p+k)^{\sigma }g_{+}+q^{\sigma }g_{-}%
\right] -(\varepsilon ^{\ast }\cdot q)\epsilon _{\alpha \beta \rho \sigma
}(p+k)^{\rho }q^{\sigma }h  \notag \\
&&-i\left[ (p+k)_{\alpha }\varepsilon _{\beta \rho \sigma \tau }\varepsilon
^{\ast \rho }(p+k)^{\sigma }q^{\tau }-\alpha \leftrightarrow \beta \right]
h_{1}  \label{16}
\end{eqnarray}%
Making use of the Dirac identity
\begin{equation}
\sigma ^{\mu \nu }\gamma ^{5}=-\frac{i}{2}\epsilon ^{\mu \nu \alpha \beta
}\sigma _{\alpha \beta }  \label{17}
\end{equation}%
in Eq.(\ref{16}), we get
\begin{eqnarray}
\left\langle D_{s}^{\ast }(k,\varepsilon )\left\vert \bar{s}i\sigma _{\mu
\nu }q^{\nu }\gamma ^{5}b\right\vert B_{c}(p)\right\rangle &=&\varepsilon
_{\mu }^{\ast }\left[ (M_{B_{c}^{-}}^{2}-M_{D_{s}^{\ast
-}}^{2})g_{+}+q^{2}g_{-}\right]  \notag \\
&&-q\cdot \varepsilon ^{\ast }\left[ q^{2}(p+k)_{\mu }g_{+}-q_{\mu }g_{-}%
\right]  \notag \\
&&+q\cdot \varepsilon ^{\ast }\left[ q^{2}(p+k)_{\mu
}-(M_{B_{c}^{-}}^{2}-M_{D_{s}^{\ast -}}^{2})q_{\mu }\right] h  \label{18}
\end{eqnarray}%
On comparing coefficents of $q_{\mu },\varepsilon _{\mu }^{\ast }$ and $%
\epsilon _{\mu \nu \alpha \beta }$ from Eqs.(\ref{13a}), (\ref{13b}), (\ref%
{16}) and (\ref{18}), we have

\begin{eqnarray}
F_{1}(q^{2}) &=&\left[ g_{+}(q^{2})-q^{2}h_{1}(q^{2})\right]  \label{19a} \\
F_{2}(q^{2}) &=&g_{+}(q^{2})+\frac{q^{2}}{M_{B_{c}^{-}}^{2}-M_{D_{s}^{\ast
-}}^{2}}g_{-}(q^{2})  \label{19b} \\
F_{3}(q^{2}) &=&-g_{-}(q^{2})-(M_{B_{c}^{-}}^{2}-M_{D_{s}^{\ast
-}}^{2})h(q^{2})  \label{19c}
\end{eqnarray}%
One can see from Eq. (\ref{19a}) and Eq. (\ref{19b}) that at $q^{2}=0,$ $%
F_{1}(0)=F_{2}(0).$ The form factors $V(q^{2}),A_{1}(q^{2})$ and $%
A_{2}(q^{2})$ can be written in terms of $g_{+},g_{-}$ and $h$ as
\begin{eqnarray}
V(q^{2}) &=&\frac{M_{B_{c}^{-}}+M_{D_{s}^{\ast -}}}{m_{b}+m_{s}}\left[
g_{+}(q^{2})-q^{2}h_{1}(q^{2})\right]  \label{19} \\
A_{1}(q^{2}) &=&\frac{M_{B_{c}^{-}}+M_{D_{s}^{\ast -}}}{m_{b}-m_{s}}\left[
g_{+}(q^{2})+\frac{q^{2}}{M_{B_{c}^{-}}^{2}-M_{D_{s}^{\ast -}}^{2}}%
g_{-}(q^{2})\right]  \label{20} \\
A_{2}(q^{2}) &=&\frac{M_{B_{c}^{-}}+M_{D_{s}^{\ast -}}}{m_{b}-m_{s}}\left[
g_{+}(q^{2})-q^{2}h(q^{2})\right] -\frac{2M_{D_{s}^{\ast -}}}{%
M_{B_{c}^{-}}-M_{D_{s}^{\ast -}}}A_{0}(q^{2})  \label{21}
\end{eqnarray}%
By looking at Eq. (\ref{19}) and Eq. (\ref{20}) it is clear that the
normalization of the form factors $V$ and $A_{1}$ at $q^{2}=0$ is determined
by a single constant $g_{+}(0),$ where as from Eq. (\ref{21}) the form
factor $A_{2}$ at $q^{2}=0$ is determined by two constants i.e. $g_{+}(0)$
and \thinspace $A_{0}(0).$

\subsection{Pole Contribution}

In $B_{c}\rightarrow D_{s}^{\ast }l^{+}l^{-}$ decay, there will be a pole
contribution to $h_{1},g_{-},h$ and $A_{0}$ from $B_{s}^{\ast
}(1^{-}),B_{sA}^{\ast }(1^{+})$ and $B_{s}(0^{-})$ mesons which can be
parameterized as
\begin{eqnarray}
h_{1}|_{pole} &=&-\frac{1}{2}\frac{g_{B_{s}^{\ast }B_{c}D_{s}^{\ast }}}{%
M_{B_{s}^{\ast }}^{2}}\frac{f_{T}^{B^{\ast }}}{1-q^{2}/M_{B^{\ast }}^{2}}=%
\frac{R_{V}}{M_{B_{s}^{\ast }}^{2}}\frac{1}{1-q^{2}/M_{B_{s}^{\ast }}^{2}}
\label{22} \\
g_{-}|_{pole} &=&-\frac{g_{B_{sA}^{\ast }B_{c}D_{s}^{\ast }}}{%
M_{B_{sA}^{\ast }}^{2}}\frac{f_{T}^{B_{sA}^{\ast }}}{1-q^{2}/M_{B_{sA}^{\ast
}}^{2}}=\frac{R_{A}^{S}}{M_{B_{sA}^{\ast }}^{2}}\frac{1}{1-q^{2}/M_{B_{sA}^{%
\ast }}^{2}}  \label{23} \\
h|_{pole} &=&\frac{1}{2}\frac{f_{B_{SA}^{\ast }B_{c}D_{s}^{\ast }}}{%
M_{B_{sA}^{\ast }}^{2}}\frac{f_{T}^{B_{SA}^{\ast }}}{1-q^{2}/M_{B_{sA}^{\ast
}}^{2}}=\frac{R_{A}^{D}}{M_{B_{sA}^{\ast }}^{2}}\frac{1}{1-q^{2}/M_{B_{sA}^{%
\ast }}^{2}}  \label{24} \\
A_{0}(q^{2})|_{pole} &=&\frac{g_{B_{s}^{\ast }B_{c}D_{s}^{\ast }}}{%
M_{B_{s}^{\ast }}^{2}}f_{B_{s}}\frac{q^{2}/M_{B}^{2}}{1-q^{2}/M_{B}^{2}}%
=R_{0}\frac{q^{2}/M_{B_{s}}^{2}}{1-q^{2}/M_{B_{s}}^{2}}  \label{25}
\end{eqnarray}%
where the quantities $R_{V},R_{A}^{S},R_{A}^{D}$ and $R_{0}$ are related to
the coupling constants $g_{B_{s}^{\ast }B_{c}D_{s}^{\ast }},g_{B_{sA}^{\ast
}B_{c}D_{s}^{\ast }}$ and $g_{B_{sA}^{\ast }B_{c}D_{s}^{\ast }}$,
respectively. Here we would like to mention that the above mentioned
couplings aries as the lower pole mass, because the higher pole masses of $%
B_{c}$ meson do not contribute for the $B_{c}\to D_s^{\ast}l^{+}l^{-}$
decay. The form factors $A_{1}(q^{2}),A_{2}(q^{2})$ and $V(q^{2})$ can be
written in terms of these quantities as
\begin{eqnarray}
V(q^{2}) &=&\frac{M_{B_{c}^{-}}+M_{D_{s}^{\ast }}}{m_{b}+m_{s}}\left[
g_{+}(q^{2})-\frac{R_{V}}{M_{B_{s}^{\ast }}^{2}}\frac{q^{2}}{%
1-q^{2}/M_{B_{s}^{\ast }}^{2}}\right]  \label{26} \\
A_{1}(q^{2}) &=&\frac{M_{B_{c}^{-}}-M_{D_{s}^{\ast -}}}{m_{b}-m_{s}}\left[
g_{+}(q^{2})+\frac{q^{2}}{M_{B_{c}^{-}}^{2}-M_{D_{s}^{\ast -}}^{2}}\tilde{g}%
_{-}(q^{2})+\frac{R_{A}^{S}}{M_{B_{sA}^{\ast }}^{2}}\frac{q^{2}}{%
1-q^{2}/M_{B_{sA}^{\ast }}^{2}}\right]  \label{27} \\
A_{2}(q^{2}) &=&\frac{M_{B_{c}^{-}}+M_{D_{s}^{\ast -}}}{m_{b}-m_{s}}\left[
g_{+}(q^{2})-\frac{R_{A}^{D}}{M_{B_{As}^{\ast }}^{2}}\frac{q^{2}}{%
1-q^{2}/M_{B_{sA}^{\ast }}^{2}}\right] -\frac{2M_{D_{s}^{\ast -}}}{%
M_{B_{c}}-M_{D_{s}^{\ast -}}}A_{0}(q^{2})  \label{28}
\end{eqnarray}%
Now, the behavior of $g_{+}(q^{2}),\tilde{g}_{-}(q^{2})$ $\ $and $%
A_{0}(q^{2})$ is known from LEET and their form is \cite{14}%
\begin{eqnarray}
g_{+}(q^{2}) &=&\frac{\xi _{\bot }(0)}{(1-q^{2}/M_{B}^{2})^{2}}=-\tilde{g}%
_{-}(q^{2})  \label{29} \\
A_{0}(q^{2}) &=&\left( 1-\frac{M_{D_{s}^{\ast -}}^{2}}{M_{B_{c}}E_{D_{s}^{%
\ast -}}}\right) \xi _{\Vert }(0)+\frac{M_{D_{s}^{\ast -}}}{M_{B_{c}}}\xi
_{\perp }(0)  \label{31} \\
E_{D_{s}^{\ast }} &=&\frac{M_{B_{c}}}{2}\left( 1-\frac{q^{2}}{M_{B_{c}}^{2}}+%
\frac{M_{D_{s}^{\ast }}^{2}}{M_{B_{c}}^{2}}\right)  \label{31a} \\
g_{+}(0) &=&\xi _{\bot }(0)  \label{32}
\end{eqnarray}%
The pole terms given in Eqs.(\ref{26}-\ref{28}) dominate near $%
q^{2}=M_{B_{s}^{\ast }}^{2}$ and $q^{2}=M_{B_{sA}^{\ast }}^{2}$. Just to
make a remark that relations obtained from the Ward identities can not be
expected to hold for the whole $q^{2}.$ Therefore, near $q^{2}=0$ and near
the pole following parametrization is suggested \cite{14}
\begin{equation}
F(q^{2})=\frac{F(0)}{\left( 1-q^{2}/M^{2}\right) (1-q^{2}/M^{\prime 2})}
\label{33}
\end{equation}%
where $M^{2}$ is $M_{B_{s}^{\ast }}^{2}$ or $M_{B_{sA}^{\ast }}^{2}$, and $%
M^{\prime }$ is the radial excitation of $M.$ The parametrization given in
Eq. (\ref{33}) not only takes into account the corrections to single pole
dominance suggested by the dispersion relation approach \cite{37, 38, 39}
but also give the correction of off-mass shell-ness of the couplings of $%
B_{s}^{\ast }$ and $B_{sA}^{\ast }$ with the $B_{c}D_{s}^{\ast }$ channel.

Since $g_{+}(0)$ and $\tilde{g}_{-}(q^{2})$ have no pole at $\
q^{2}=M_{B_{s}^{\ast }}^{2},$ hence we get
\begin{equation*}
V(q^{2})(1-\frac{q^{2}}{M_{B^{\ast }}^{2}})|_{q^{2}=M_{B^{\ast
}}^{2}}=-R_{V}\left( \frac{M_{B_{c}}+M_{D_{s}^{\ast }}}{m_{b}-m_{s}}\right)
\end{equation*}%
This becomes
\begin{equation}
R_{V}\equiv -\frac{1}{2}g_{B_{s}^{\ast }B_{c}D_{s}^{\ast }}f_{B^{\ast}_s}=-%
\frac{g_{+}(0)}{1-M_{B^{\ast }}^{2}/M_{B^{\ast }}^{\prime 2}}  \label{34}
\end{equation}%
and similarly%
\begin{equation}
R_{A}^{D}\equiv \frac{1}{2}f_{B_{sA}^{\ast }B_{c}D_{s}^{\ast
}}f^{B^{\ast}_{sA}}_{T}=-\frac{g_{+}(0)}{1-M_{B_{sA}^{\ast
}}^{2}/M_{B_{sA}^{\ast }}^{\prime 2}}  \label{35}
\end{equation}
We cannot use the parametrization given in Eq.(\ref{33}) for the form factor
$A_{1}(q^{2}),$ since near $q^{2}=0,$ the behavior of $A_{1}(q^{2})$ is $%
g_{+}(q^{2})\left[ 1-q^{2}/\left( M_{B_{c}^{-}}^{2}-M_{D_{s}^{\ast
-}}^{2}\right) \right] ,$ therefore we can write $A_{1}(q^{2})$ as follows%
\begin{equation}
A_{1}(q^{2})=\frac{g_{+}(0)}{\left( 1-q^{2}/M_{B_{sA}^{\ast }}^{2}\right)
\left( 1-q^{2}/M_{B_{sA}^{\ast }}^{\prime 2}\right) }\left( 1-\frac{q^{2}}{%
M_{B_{c}^{-}}^{2}-M_{D_{s}^{\ast }}^{2}}\right)  \label{36}
\end{equation}%
The only unkonown parameter in the above form factors calculation is $%
g_{+}(0)$ and its value can be extracted by using the central value of
branching ratio for the decay $B_{c}^{-}\rightarrow D_{s}^{\ast -}\gamma $
\cite{41}. From the formula of decay rate
\begin{equation}
\Gamma \left( B_{c}\rightarrow D_{s}^{\ast }\gamma \right) =\frac{%
G_{F}^{2}\alpha }{32\pi ^{4}}\left\vert V_{tb}V_{ts}^{\ast }\right\vert
^{2}m_{b}^{2}M_{B_{c}}^{3}\times \left( 1-\frac{M_{D_{s}^{\ast }}^{2}}{%
M_{B_{c}}^{2}}\right) ^{3}\left\vert C_{7}^{eff}\right\vert ^{2}\left\vert
g_{+}(0)\right\vert ^{2}  \label{36a}
\end{equation}%
and by putting the values of everything one can find the value of unknown
parameter $g_{+}(0)=0.32\pm 0.1$. In the forthcoming analysis we use the
value of $g_{+}(0)=0.42$ which was calculated in ref. \cite{41}.

Using $f_{B_{c}}=0.35$ GeV we have prediction from Eq.(\ref{34}) that%
\begin{equation}
g_{B_{s}^{\ast }B_{c}D_{s}^{\ast }}=10.38 GeV^{-1}.  \label{38}
\end{equation}%
Similarly the ratio of $S$ and $D$ wave couplings are predicted to be
\begin{equation}
\frac{g_{B_{sA}^{\ast }B_{c}D_{s}^{\ast }}}{f_{B_{sA}^{\ast
}B_{c}D_{s}^{\ast }}}=-0.42 GeV^{2}  \label{38A}
\end{equation}%
The different values of the $F(0)$ are
\begin{eqnarray}
V(0) &=&\frac{M_{B_{c}^{-}}+M_{D_{s}^{\ast -}}}{m_{b}+m_{s}}g_{+}(0)
\label{38a} \\
A_{1}(0) &=&\frac{M_{B_{c}^{-}}-M_{D_{s}^{\ast -}}}{m_{b}-m_{s}}g_{+}(0)
\label{39} \\
A_{2}(0) &=&\frac{M_{B_{c}^{-}}+M_{D_{s}^{\ast -}}}{m_{b}-m_{s}}g_{+}(0)-%
\frac{2M_{D_{s}^{\ast -}}}{M_{B_{c}^{-}}-M_{D_{s}^{\ast -}}}A_{0}(0)
\label{40}
\end{eqnarray}%
The calculation of the numerical values of $V(0)$ and $A_{1}(0)$ is quite
trivial but for the value of $A_{2}(0),$ the value of $A_{0}(0)$ has to be
known. Although LEET does not give any relationship between $\xi _{||}(0)$
and $\xi _{\perp }(0)$, but in LCSR $\xi _{||}(0)$ and $\xi _{\perp }(0)$
are related due to numerical coincidence \cite{42}
\begin{equation}
\xi _{||}(0)\simeq \xi _{\perp }(0)=g_{+}(0)  \label{41}
\end{equation}%
From Eq. (\ref{31}) we have
\begin{equation*}
A_{0}(0)=1.12g_{+}(0)
\end{equation*}

The value of the form factors at $q^{2}=0$ is given in Table-1%
\begin{table}[tb]
\caption{Values of the form factors at $q^2=0$.}
\label{Table I}
\begin{center}
\begin{tabular}[t]{cccccc}
\hline\hline
$V(0)$ & $A_{1}(0)$ & $\tilde{A}_{2}(0)$ & $A_{0}(0)$  &  &  \\ \hline\hline
$0.51\pm 0.17$ & $0.28\pm 0.08$ & $0.22\pm 0.07$ & $0.35\pm 0.11$ &  &  \\
\hline\hline
&  &  &  &  &
\end{tabular}%
\end{center}
\end{table}
and can be extrapolated for the other values of $q^{2}$ as follows:
\begin{eqnarray}
V(q^{2}) &=&\frac{V(0)}{(1-q^{2}/M_{B_{s}^{\ast
}}^{2})(1-q^{2}/M_{B_{s}^{\ast }}^{\prime 2})}  \label{42} \\
A_{1}(q^{2}) &=&\frac{A_{1}(0)}{(1-q^{2}/M_{B_{sA}^{\ast
}}^{2})(1-q^{2}/M_{B_{sA}^{\ast }}^{\prime 2})}  \label{43} \\
A_{2}(q^{2}) &=&\frac{\tilde{A}_{2}(0)}{(1-q^{2}/M_{B_{sA}^{\ast
}}^{2})(1-q^{2}/M_{B_{sA}^{\ast }}^{\prime 2})}  \notag \\
&&-\frac{2M_{D_{s}^{\ast -}}}{M_{B_{c}^{-}}-M_{D_{s}^{\ast -}}}\frac{A_{0}(0)%
}{(1-q^{2}/M_{B_{s}}^{2})(1-q^{2}/M_{B_{s}}^{\prime 2})}  \notag \\
&&  \label{44}
\end{eqnarray}%
The behavior of form factors $V(q^{2}),$ $A_{1}(q^{2})$ and $A_{2}(q^{2})$
are shown in Fig. 1.
\begin{figure}[h]
\begin{center}
\begin{tabular}{cccc}
\vspace{-0.5cm} \includegraphics[scale=0.6]{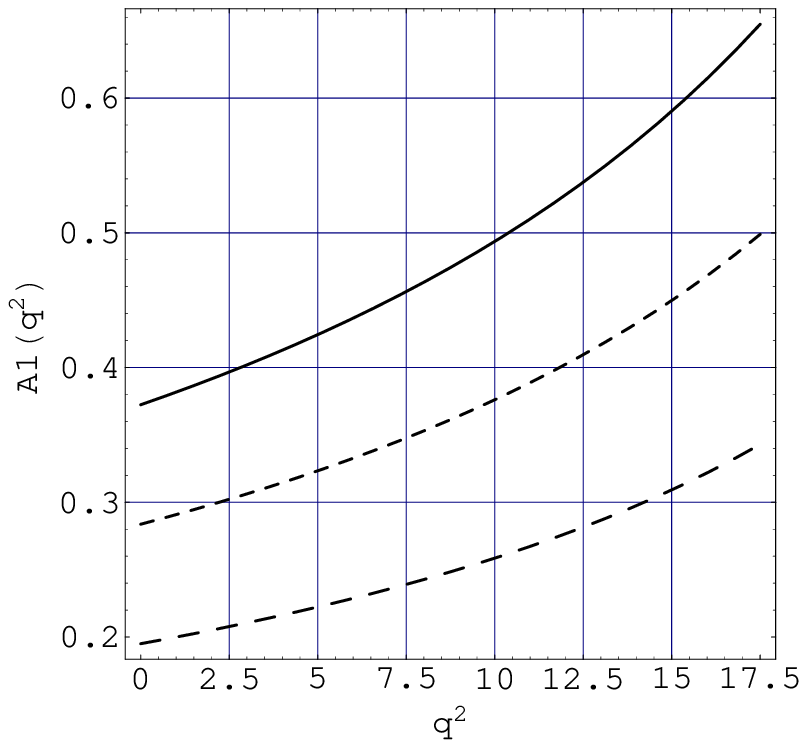} %
\includegraphics[scale=0.6]{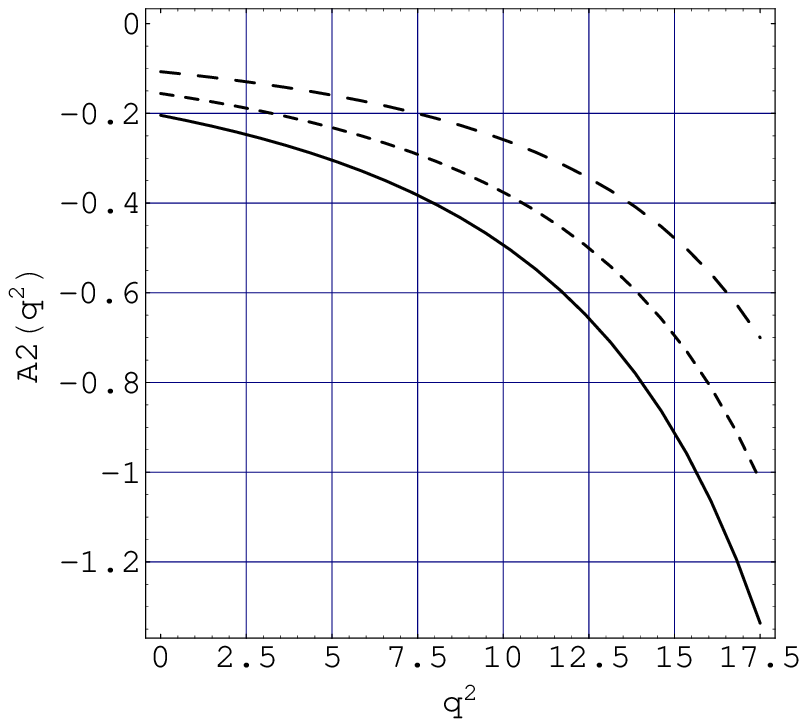} \includegraphics[scale=0.6]{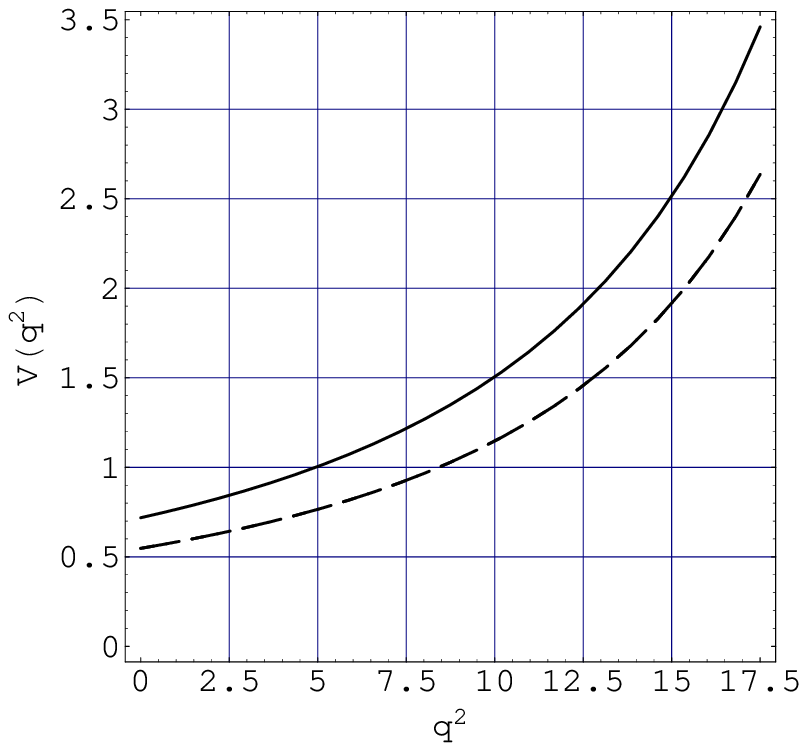}
\put (-350,160){(a)} \put
(-220,160){(b)} \put (-80,160){(c)}&  &
\end{tabular}%
\end{center}
\caption{Form factors are plotted as a function of $q^2$. Solid line,
dashed line and long-dashed line correspond to $g_{+}(0)$ equal to 0.42, 0.32 and
0.22 respectively.}
\label{Form Factors}
\end{figure}

\section{Physical Observables for $B_{c}\rightarrow D_{s}^{\ast }l^{+}l^{-}$}

In this section we will present the calculations of the physical observables
like the decay rates and the helicity fractions of $D_{s}^{\ast }$ meson.
From Eq. (\ref{5a}) it is straightforward to write
\begin{eqnarray}
\mathcal{M}_{B_{c}\rightarrow D_{s}^{\ast }l^{+}l^{-}}&=& -\frac{G_{F}\alpha
}{2\sqrt{2}\pi }V_{tb}V_{ts}^{\ast }\left[ T_{\mu }^{1}(\bar{l}\gamma ^{\mu
}l)+T_{\mu }^{2}\left( \bar{l}\gamma ^{\mu }\gamma ^{5}l\right) \right]
\label{59}
\end{eqnarray}%
where%
\begin{eqnarray}
T_{\mu }^{1} &=&f_{1}(q^{2})\epsilon _{\mu \nu \alpha \beta }\varepsilon
^{\ast \nu }p^{\alpha }k^{\beta }+if_{2}(q^{2})\varepsilon _{\mu }^{\ast
}+if_{3}(q^{2})(\varepsilon ^{\ast }\cdot q)P_{\mu }  \label{60} \\
T_{\mu }^{2} &=&f_{4}(q^{2})\epsilon _{\mu \nu \alpha \beta }\varepsilon
^{\ast \nu }p^{\alpha }k^{\beta }+if_{5}(q^{2})\varepsilon _{\mu }^{\ast
}+if_{6}(q^{2})(\varepsilon ^{\ast }\cdot q)P_{\mu }  \label{61}
\end{eqnarray}

The functions $f_{1}$ to $f_{6}$ in Eq.(\ref{60}) and Eq. (\ref{61}) are
known as auxiliary functions, which contains both long distance (Form
factors) and short distance (Wilson coefficients) effects and these can be
written as

\begin{eqnarray}
f_{1}(q^{2}) &=&4(m_{b}+m_{s})\frac{C_{7}^{eff}}{q^{2}}%
F_{1}(q^{2})+C_{9}^{eff}\frac{V(q^{2})}{M_{B_{c}}+M_{D_{s}^{\ast }}}  \notag
\\
f_{2}(q^{2}) &=&\frac{C_{7}^{eff}}{q^{2}}4(m_{b}-m_{s})F_{2}(q^{2})\left(
M_{B_{c}}^{2}-M_{D_{s}^{\ast }}^{2}\right) +C_{9}^{eff}A_{1}(q^{2})\left(
M_{B_{c}}+M_{D^{\ast }}\right)  \notag \\
f_{3}(q^{2}) &=&-\left[ C_{7}^{eff}4(m_{b}-m_{s})\left( F_{2}(q^{2})+q^{2}%
\frac{F_{3}(q^{2})}{\left( M_{B_{c}}^{2}-M_{D_{s}^{\ast }}^{2}\right) }%
\right) +C_{9}^{ff}\frac{A_{2}(q^{2})}{M_{B_{c}}+M_{D_{s}^{\ast }}}\right]
\notag \\
f_{4}(q^{2}) &=&C_{10}\frac{V(q^{2})}{M_{B_{c}}+M_{D_{s}^{\ast }}}  \notag \\
f_{5}(q^{2}) &=&C_{10}A_{1}(q^{2})\left( M_{B_{c}}+M_{D_{s}^{\ast }}\right)
\notag \\
f_{6}(q^{2}) &=&-C_{10}\frac{A_{2}(q^{2})}{M_{B_{c}}+M_{D_{s}^{\ast }}}
\notag \\
f_{0}(q^{2}) &=&C_{10}A_{0}(q^{2})  \label{62}
\end{eqnarray}%
The next task is to calculate the decay rate and the helicity fractions of $%
D_{s}^{\ast }$ meson in terms of these auxiliary functions.

\subsection{The Differential Decay Rate of $B_{c}\rightarrow D_{s}^{\ast
}l^{+}l^{-}$}

In the rest frame of $B_{c}$ meson the differential decay width of $%
B_{c}\rightarrow D_{s}^{\ast }l^{+}l^{-}$ can be written as
\begin{eqnarray}
\frac{d\Gamma (B_{c}\rightarrow D_{s}^{\ast }l^{+}l^{-})}{dq^{2}}&=&\frac{1}{%
\left( 2\pi \right) ^{3}}\frac{1}{32M_{B_{c}}^{3}}%
\int_{-u(q^{2})}^{+u(q^{2})}du\left\vert \mathcal{M}_{B_{c}\rightarrow
D_{s}^{\ast }l^{+}l^{-}}\right\vert ^{2}  \label{62a}
\end{eqnarray}%
where
\begin{eqnarray}
q^{2} &=&(p_{l^{+}}+p_{l^{-}})^{2}  \label{62b} \\
u &=&\left( p-p_{l^{-}}\right) ^{2}-\left( p-p_{l^{+}}\right) ^{2}
\label{62c}
\end{eqnarray}%
Now the limits on $q^{2}$ and $u$ are
\begin{eqnarray}
4m_{l}^{2} &\leq &q^{2}\leq (M_{B_{c}}-M_{D_{s}^{\ast }})^{2}  \label{62d} \\
-u(q^{2}) &\leq &u\leq u(q^{2})  \label{62e}
\end{eqnarray}%
with%
\begin{equation}
u(q^{2})=\sqrt{\lambda \left( 1-\frac{4m_{l}^{2}}{q^{2}}\right) }
\label{62f}
\end{equation}%
where%
\begin{equation*}
\lambda \equiv \lambda (M_{B_{c}}^{2},M_{D_{s}^{\ast
}}^{2},q^{2})=M_{B_{c}}^{4}+M_{D_{s}^{\ast
}}^{4}+q^{4}-2M_{B_{c}}^{2}M_{D_{s}^{\ast }}^{2}-2M_{D_{s}^{\ast
}}^{2}q^{2}-2q^{2}M_{B_{c}}^{2}
\end{equation*}%
The decay rate of $B_{c}\rightarrow D_{s}^{\ast }l^{+}l^{-}$ can easily
obtained in terms of auxiliary function by integrating on $u$ (c.f. Eq. (\ref%
{62a})) as
\begin{eqnarray}  \label{63a}
\frac{d\Gamma (B_{c}\rightarrow D_{s}^{\ast }l^{+}l^{-})}{dq^{2}} &=& \frac{
G_{F}^{2}\left\vert V_{tb}V_{ts}^{\ast }\right\vert ^{2}\alpha ^{2}}{
2^{11}\pi ^{5}3M_{B_{c}}^{3}M_{D_{s}^{\ast }}^{2}q^{2}}u(q^{2})\bigg[%
24\left\vert f_{0}(q^{2})\right\vert ^{2}m_{l}^{2}M_{D_{s}^{\ast}}^{2}\lambda
\notag \\
&&+8M_{D_{s}^{\ast }}^{2}q^{2}\lambda[ (2m_{l}^{2}+q^{2})\left\vert
f_{1}(q^{2})\right\vert ^{2}-(4m_{l}^{2}-q^{2})\left\vert
f_{4}(q^{2})\right\vert ^{2}]  \notag \\
&&+\lambda [(2m_{l}^{2}+q^{2})\left\vert f_{2}(q^{2})
+(M_{B_{c}}^{2}-M_{D_{s}^{\ast }}^{2}-q^{2})f_{3}(q^{2})\right\vert ^{2}
\notag \\
&&-(4m_{l}^{2}-q^{2}) \left\vert f_{5}(q^{2})+(M_{B_{c}}^{2}-M_{D_{s}^{\ast
}}^{2}-q^{2})f_{6}(q^{2})\right\vert ^{2}]  \notag \\
&&+4M_{D_{s}^{\ast }}^{2}q^{2}[ (2m_{l}^{2}+q^{2})\left( 3\left\vert
f_{2}(q^{2})\right\vert ^{2} -\lambda \left\vert f_{3}(q^{2})\right\vert
^{2}\right)  \notag \\
&&-(4m_{l}^{2}-q^{2})\left( 3\left\vert f_{5}(q^{2})\right\vert ^{2}-\lambda
\left\vert f_{6}(q^{2})\right\vert ^{2}\right)]\bigg]  \notag \\
\end{eqnarray}

\subsection{HELICITY FRACTIONS OF $D_{s}^{\ast }$ IN $B_{c}\rightarrow
D_{s}^{\ast }l^{+}l^{-}$}

We now discuss helicity fractions of $D_{s}^{\ast }$ in $B_{c}\rightarrow
D_{s}^{\ast }l^{+}l^{-}$ which are intersting variable and are as such
independent of the uncertainities arising due to form factors and other
input parameters. The final state meson helicity fractions were already
discussed in literature for $B\rightarrow K^{\ast }\left( K_{1}\right)
l^{+}l^{-}$ decays \cite{22, 23}. Even for the \ $K^{\ast }$ vector meson,
the longitudinal helicity fraction $f_{L}$ has been measured by Babar
collaboration for the decay $B\rightarrow K^{\ast }l^{+}l^{-}(l=e,\mu )$ in
two bins of momentum transfer and the results are \cite{46}
\begin{eqnarray}
f_{L} &=&0.77_{-0.30}^{+0.63}\pm 0.07 , \ \ \ \ \ 0.1\leq q^{2}\leq 8.41
GeV^{2}  \notag \\
&&  \label{64} \\
f_{L} &=&0.51_{-0.25}^{+0.22}\pm 0.08 , \ \ \ \ \ q^{2}\geq 10.24GeV^{2}
\notag
\end{eqnarray}%
while the average value of $f_{L}$ in full $q^{2}$ range is
\begin{equation}
f_{L}=0.63_{-0.19}^{+0.18}\pm 0.05 , \ \ q^{2}\geq 0.1 GeV^{2}  \label{65}
\end{equation}

The explicit expression of the helicity fractions for $B_{c}^{-}\rightarrow
D_{s}^{\ast -}l^{+}l^{-}$ decay can be written as%
\begin{eqnarray}
\frac{d\Gamma _{L}(q^{2})}{dq^{2}} &=&\frac{G_{F}^{2}\left\vert
V_{tb}V_{ts}^{\ast }\right\vert ^{2}\alpha ^{2}}{2^{11}\pi ^{5}}\frac{%
u(q^{2})}{M_{B_{c}}^{3}}\times  \notag \\
&&\frac{1}{3}\frac{1}{q^{2}M_{D_{s}^{\ast }}^{2}}\left[
\begin{array}{c}
24\left\vert f_{0}(q^{2})\right\vert ^{2}m_{l}^{2}M_{D_{s}^{\ast
}}^{2}\lambda +(2m_{l}^{2}+q^{2})\left\vert \left(
M_{B_{c}}^{2}-M_{D_{s}^{\ast }}^{2}-q^{2}\right) f_{2}(q^{2})+\lambda
f_{3}(q^{2})\right\vert ^{2} \\
+\left( q^{2}-4m_{l}^{2}\right) \left\vert \left(
M_{B_{c}}^{2}-M_{D_{s}^{\ast }}^{2}-q^{2}\right) f_{5}(q^{2})+\lambda
f_{6}(q^{2})\right\vert ^{2}%
\end{array}%
\right]  \notag \\
&&  \label{66}
\end{eqnarray}%
\begin{eqnarray}
\frac{d\Gamma _{+}(q^{2})}{dq^{2}} &=&\frac{G_{F}^{2}\left\vert
V_{tb}V_{ts}^{\ast }\right\vert ^{2}\alpha ^{2}}{2^{11}\pi ^{5}}\frac{%
u(q^{2})}{M_{B_{c}}^{3}}\times  \notag \\
&&\frac{4}{3}\left[ \left( q^{2}-4m_{l}^{2}\right) \left\vert f_{5}(q^{2})-%
\sqrt{\lambda }f_{4}(q^{2})\right\vert ^{2}+\left( q^{2}+2m_{l}^{2}\right)
\left\vert f_{2}(q^{2})-\sqrt{\lambda }f_{1}(q^{2})\right\vert ^{2}\right]
\label{67} \\
\frac{d\Gamma _{-}(q^{2})}{dq^{2}} &=&\frac{G_{F}^{2}\left\vert
V_{tb}V_{ts}^{\ast }\right\vert ^{2}\alpha ^{2}}{2^{11}\pi ^{5}}\frac{%
u(q^{2})}{M_{B_{c}}^{3}}\times  \notag \\
&&\frac{4}{3}\left[ \left( q^{2}-4m_{l}^{2}\right) \left\vert f_{5}(q^{2})+%
\sqrt{\lambda }f_{4}(q^{2})\right\vert ^{2}+\left( q^{2}+2m_{l}^{2}\right)
\left\vert f_{2}(q^{2})+\sqrt{\lambda }f_{1}(q^{2})\right\vert ^{2}\right]
\label{68}
\end{eqnarray}%
where the auxiliary functions and the corresponding form factors are given
in Eq.(\ref{62}) and Eqs.(\ref{42}-\ref{44}). Finally the longitudinal and
transverse helicity amplitude becomes
\begin{eqnarray}
f_{L}(q^{2}) &=&\frac{d\Gamma _{L}(q^{2})/dq^{2}}{d\Gamma (q^{2})/dq^{2}}
\notag \\
f_{\pm }(q^{2}) &=&\frac{d\Gamma _{\pm }(q^{2})/dq^{2}}{d\Gamma
(q^{2})/dq^{2}}  \notag \\
f_{T}(q^{2}) &=&f_{+}(q^{2})+f_{-}(q^{2})  \label{69}
\end{eqnarray}%
so that \ the sum of the longitudinal and transverse helicity amplitudes is
equal to one i.e. $f_{L}(q^{2})+f_{T}(q^{2})=1$ for each value of $q^{2}$%
\cite{22}.

\section{Numerical Analysis.}

In this section we present the numerical analysis of the branching ratio and
helicity fractions of $D_{s}^{\ast }$ meson in $B_{c}\rightarrow D_{s}^{\ast
}l^{+}l^{-}(l=\mu ,\tau )$ both in the SM and in ACD model. One of the main
input parameters are the form factors which are non perturbative quantities
and are the major source of uncertainties. Here we calculated the form
factors using the Ward identities and their dependence on momentum transfer $%
q^{2}$ is given in Section III. We have used next-to-leading order
approximation for the Wilson Coefficients at the renormalization scale $\mu
=m_{b}.$ It has already been mentioned that besides the contribution in the $%
C_{9}^{eff}$, there are long distance contributions resulting from the $c%
\bar{c}$ resonances like $J/\psi $ and its excited states. For the present
analysis \ we do not take into account these long distance effects.

The numerical results for the decay rates and helicity fractions of $%
D_{s}^{\ast }$ for the decay mode $B_{c}\rightarrow D_{s}^{\ast }l^{+}l^{-}$
both for the SM and ACD model are depicted in Figs. 2-4. Figs. 2 (a, b)
shows the differential decay rate of $B_{c}\rightarrow D_{s}^{\ast
}l^{+}l^{-}(l=\mu ,\tau ).$ One can see that there is a significant
enhancement in the decay rate due to KK-contribution for $1/R=300$ GeV,
whereas the value of the decay rate is shifted towards the SM at large value
of $1/R$ , both in small and large value of momentum transfer $q^{2}.$

\begin{figure}[h]
\begin{center}
\begin{tabular}{ccc}
\vspace{-2cm} \includegraphics[scale=0.6]{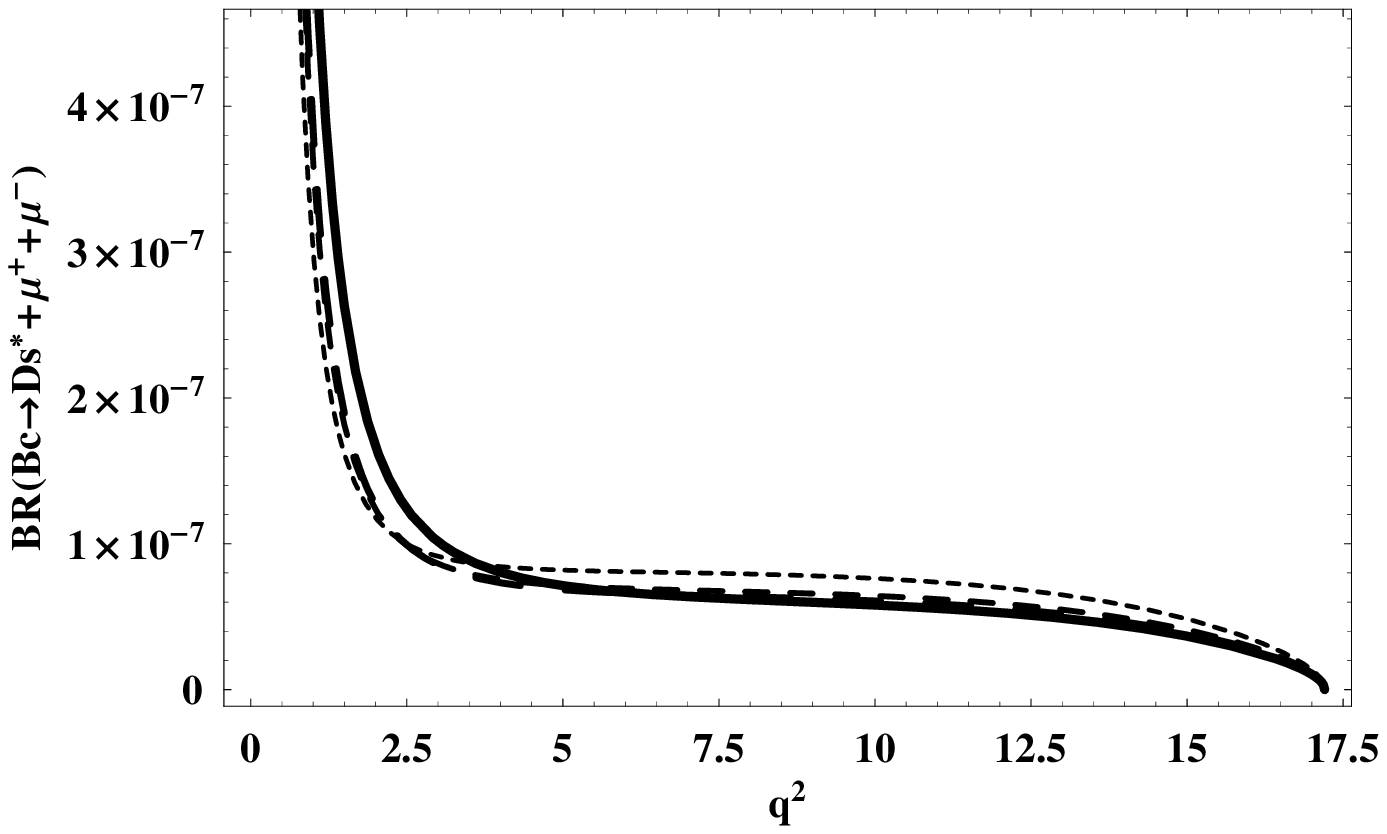} %
\includegraphics[scale=0.6]{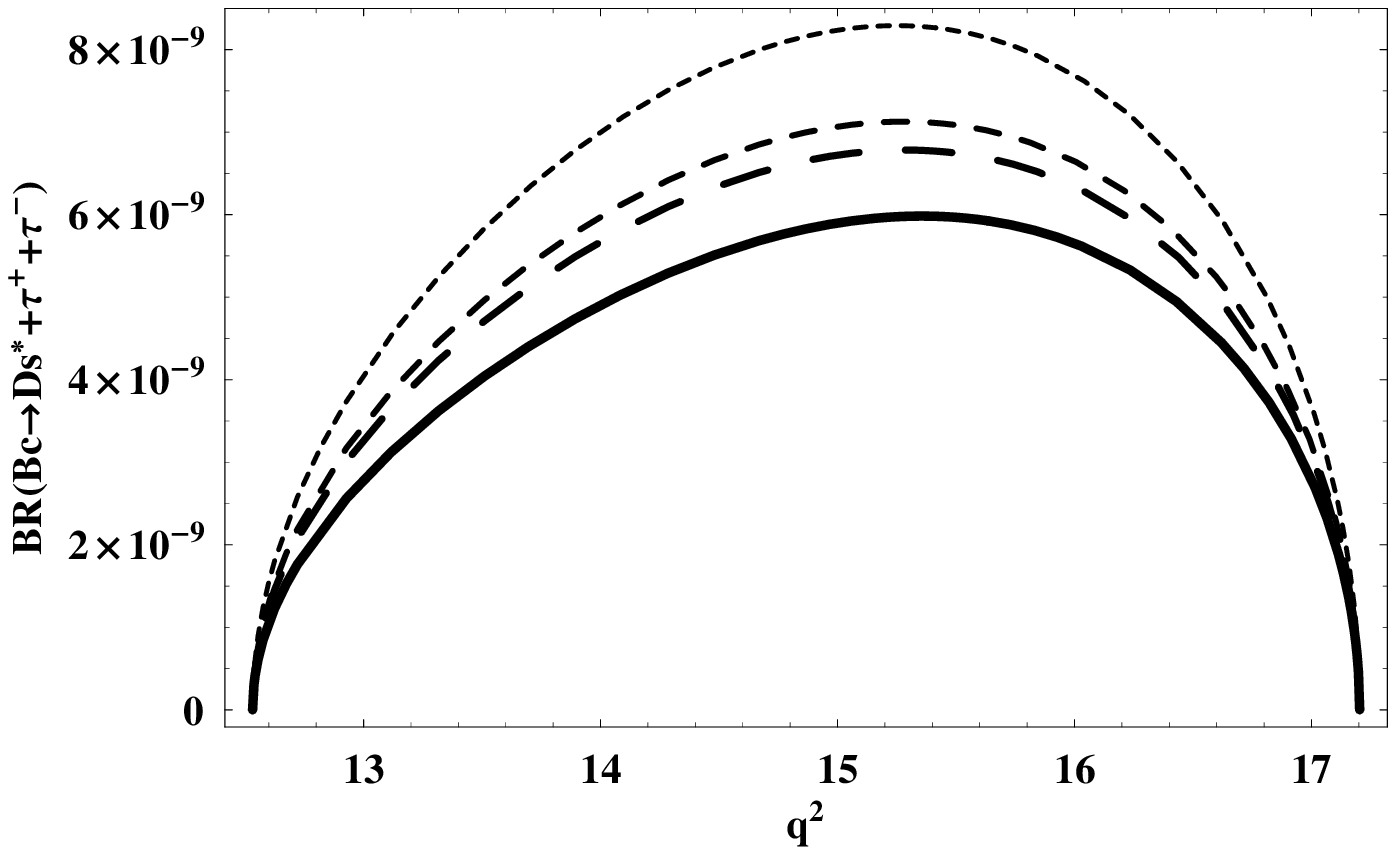} \put (-350,220){(a)} \put
(-100,220){(b)} &  &
\end{tabular}%
\end{center}
\caption{Branching ratio for the $B\rightarrow D_{s}^{\ast}l^{+}l ^{-}$ $(l=%
\protect\mu, \protect\tau)$ decays as functions of $q^{2}$ for different
values of $1/R$. Solid line correspond to SM value,dotted line is for $%
1/R=300$, dashed is for $1/R=500$, long dashed line is for $1/R=700$. }
\label{Branching Ratio}
\end{figure}

In general the sensitivity on $1/R$ is usually masked by the uncertainties
which arises due to the number of sources. Among them the major one lies in
the numerical analysis of $B_{c}\to D^{*}_s l^{+}l^{-}$ decay originated
from the $B_{c}\to D^{*}_s$ transition form factors calculated in the
present approach as shown in Table I, which can bring about almost $40\%$
errors to the differential decay rate of above mentioned decay, which showed
that it is not a very suitable tool to look for the new physics. The large
uncertainties involved in the form factors are mainly from the variations of
the decay constant of $B_c$ meson and also there are some uncertainties from
the strange quark mass $m_s$, which are expected to be very tiny on account
of the negligible role of $m_s$ suppressed by the much larger energy scale
of $m_{b}$. Moreover, the uncertainties of the charm quark and bottom quark
mass are at the $1\%$ level, which will not play significant role in the
numerical analysis and can be dropped out safely. It also needs to be
stressed that these hadronic uncertainties almost have no influence on the
various asymmetries including the polarization asymmetries of final state
meson on account of the serious cancelation among different polarization
states and this make them one of the best tool to look for physics beyond
the SM.

Figs. 3 (a, b) shows the longitudinal and transverse helicity fractions of $%
D_{s}^{\ast }$ for the decay $B_{c}\rightarrow D_{s}^{\ast }\mu ^{+}\mu ^{-}$
where we have used the central value of the form factors which we have
calculated in Section III. Choosing the different values of compactification
radius $1/R$, one can see from the graphs that the effect of extra
dimensions are quite significant at a particular region of $q^{2}$. These
effects are constructive for the case of transverse helicity fraction and
destructive for the case of longitudinal helicity fraction.

\begin{figure}[h]
\begin{center}
\begin{tabular}{ccc}
\vspace{-1cm} \includegraphics[scale=0.5]{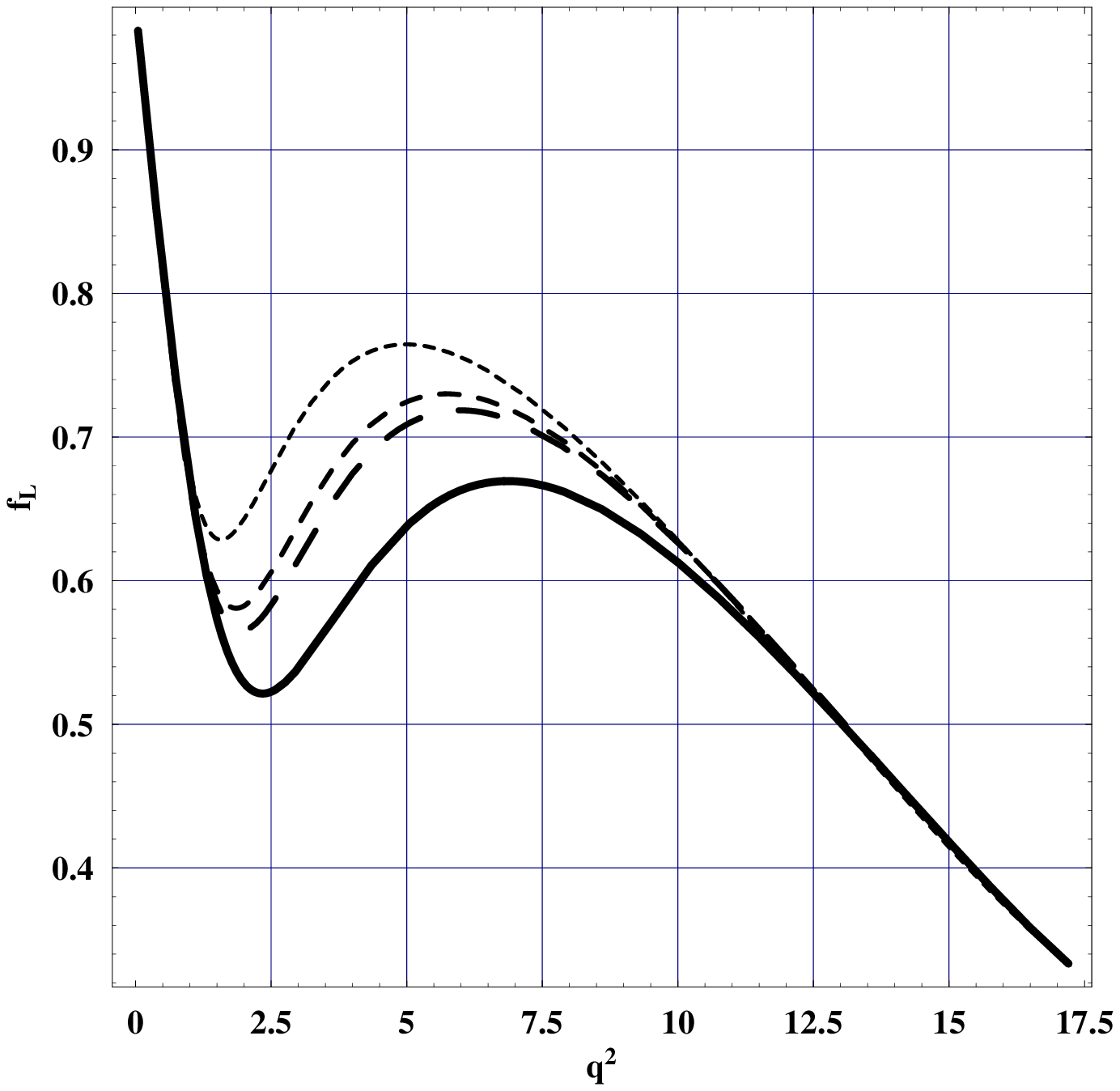} %
\includegraphics[scale=0.5]{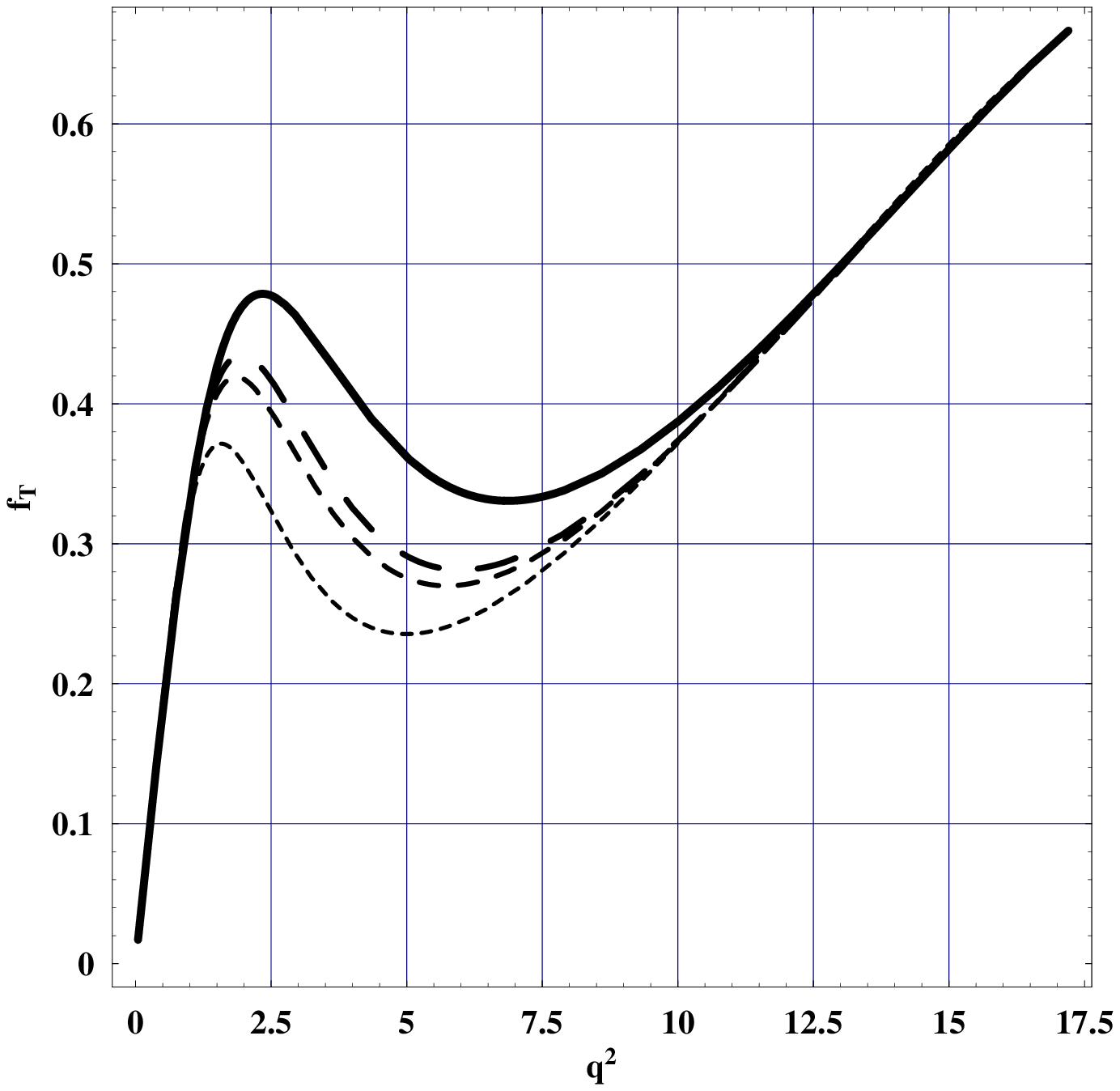} \put (-300,220){(a)} \put
(-100,220){(b)} &  &
\end{tabular}%
\end{center}
\caption{Longitudinal Lepton polarization Fig.1a and Transverse Lepton
polarization Fig.2b for the $B\rightarrow D_{s}^{\ast}\protect\mu^{+}\protect%
\mu ^{-}$ decays as functions of $q^{2}$ for different values of $1/R$.
Solid line correspond to SM value,dotted line is for $1/R=300$, dashed is
for $1/R=500$, long dashed line is for $1/R=700$. }
\label{Longitudinal Lepton Polarization for muons}
\end{figure}
Similarly, Figs. 4 (a,b) show the helicity fraction of $D_{s}^{\ast }$ for
the decay $B_{c}\rightarrow D_{s}^{\ast }\tau ^{+}\tau ^{-}$ where one can
see that the effects of the extra dimensions are mild as compared to the
case of $B_{c}\rightarrow D_{s}^{\ast }\mu ^{+}\mu ^{-}$ . Moreover from
Figs.2-4 it is clear that each value of momentum transfer $q^{2}$ the sum of
the longitudinal and transverse helicity fractions are equal to one, i.e. $%
f_{L}(q^{2})+f_{T}(q^{2})=1$.

\begin{figure}[h]
\begin{center}
\begin{tabular}{ccc}
\vspace{-1cm} \includegraphics[scale=0.5]{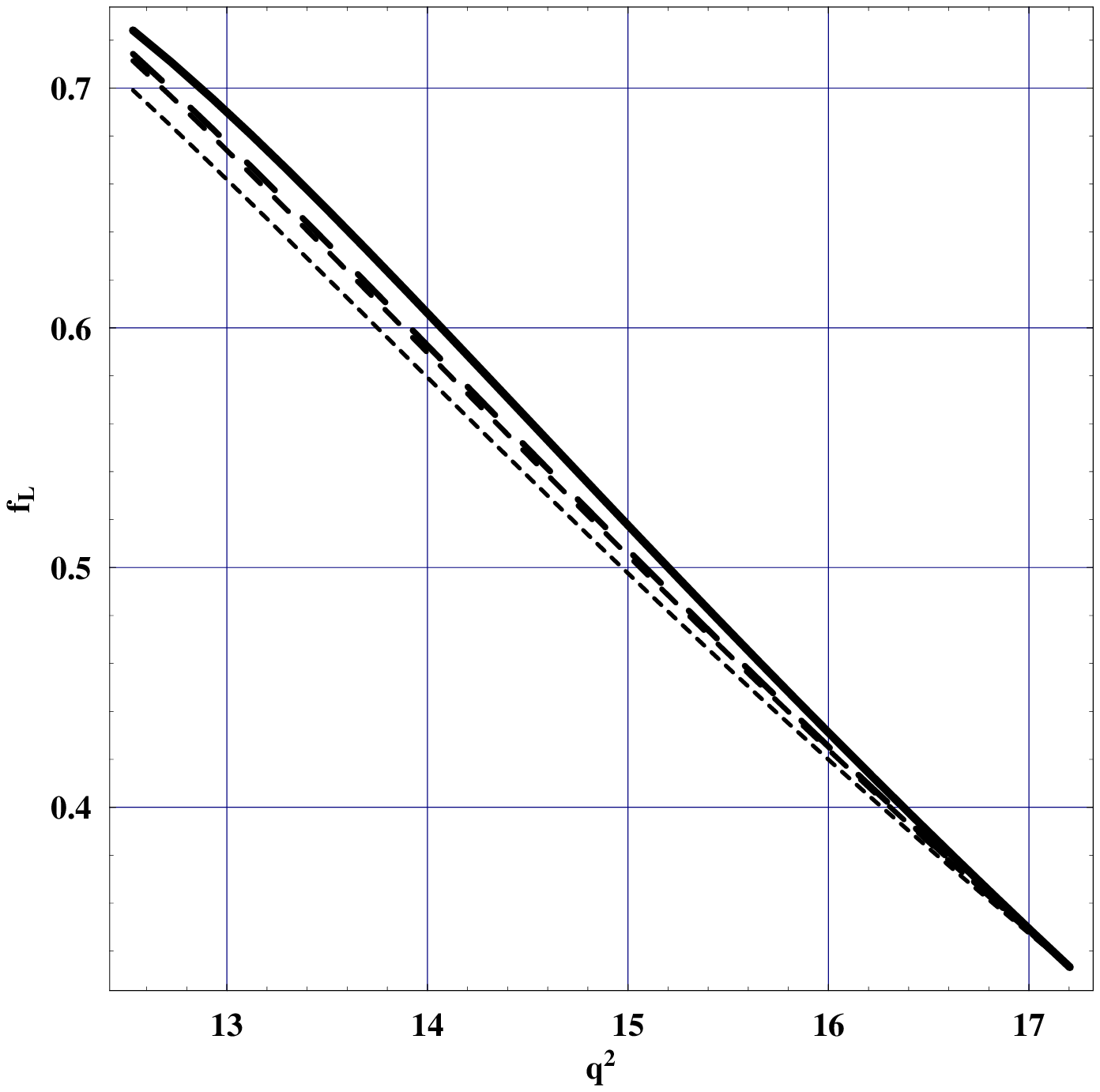} %
\includegraphics[scale=0.5]{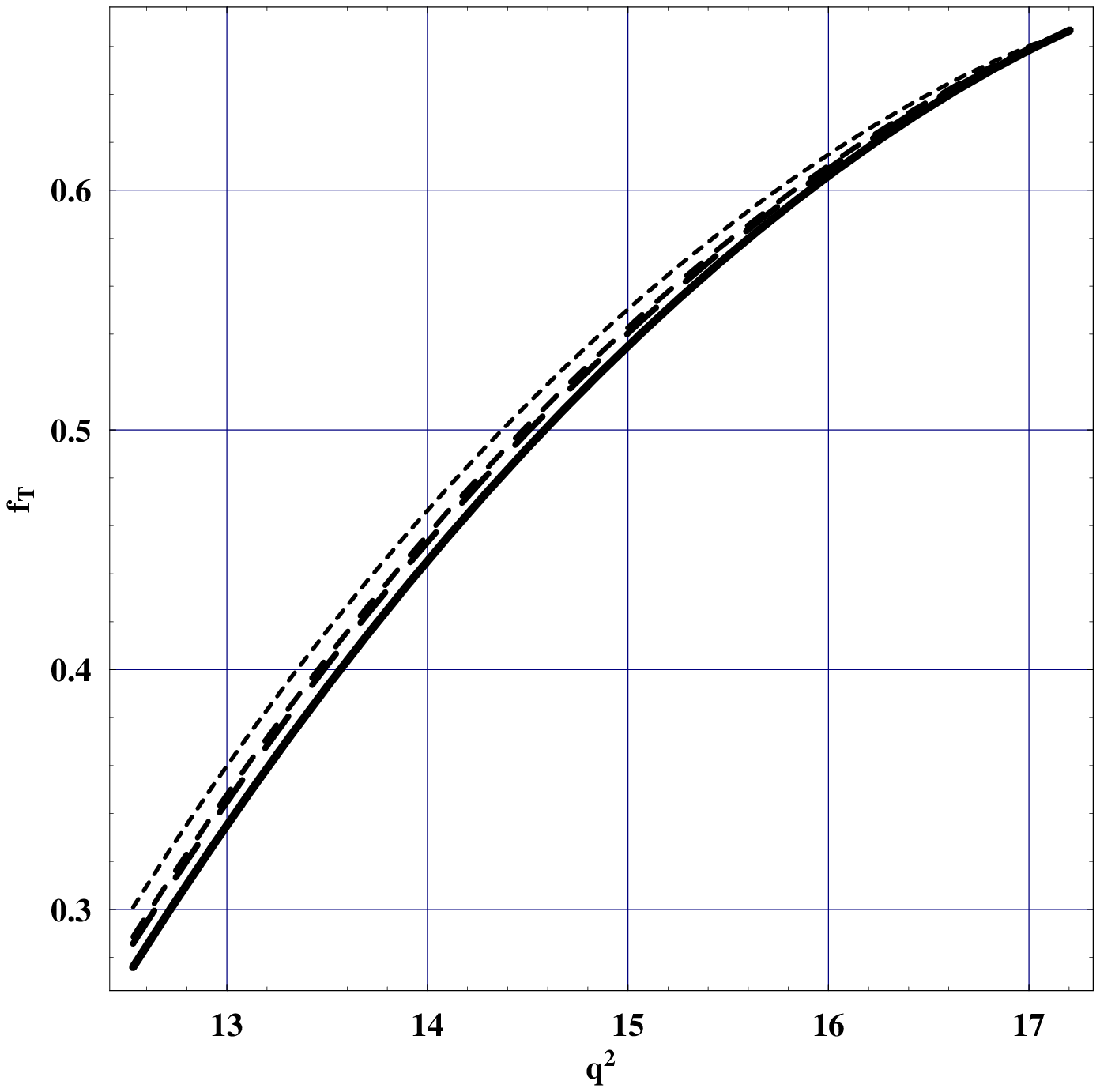} \put (-300,220){(a)} \put
(-100,220){(b)} &  &
\end{tabular}%
\end{center}
\caption{Longitudinal Lepton polarization Fig.1a and Transverse Lepton
polarization Fig.2b for the $B\rightarrow D_{s}^{\ast}\protect\tau^{+}%
\protect\tau ^{-}$ decays as functions of $q^{2}$ for different values of $%
1/R$. Solid line correspond to SM value,dotted line is for $1/R=300$, dashed
is for $1/R=500$, long dashed line is for $1/R=700$. }
\label{Longitudinal Lepton Polarization for muons}
\end{figure}

\section{Conclusion:}

We investigated the semileptonic decay $B_{c}\rightarrow D_{s}^{\ast
}l^{+}l^{-}$ $(l=\mu ,\tau )$ using the Ward identities. The form factors
have been calculated and we found that the normalization of \ the form
factors in terms of a single universal constant $g_{+}(0)$. The value of $%
g_{+}(0)=0.42$ \ is obtained from the decay $B_{c}\rightarrow D_{s}^{\ast
}\gamma $ \cite{41}. Considering the radial excitation at lower pole masses $%
M$ ( where $M=M_{B_{s}^{\ast }}$ and $M_{B_{sA}^{\ast }})$ one can predict
the coupling of $B_{s}^{\ast }$ with $B_{c}D_{s}^{\ast }$ channel as
indicated in Eq.(\ref{38}) which is $g_{B_{s}^{\ast }B_{c}D_{s}^{\ast
}}=10.38$ GeV$^{-1}.$ Also we predicted the ratio of $S$ and $D$ wave
couplings $\frac{g_{B_{sA}^{\ast }B_{c}D_{s}^{\ast }}}{f_{B_{sA}^{\ast
}B_{c}D_{s}^{\ast }}}=-0.42$ $GeV^{2}$ given in Eq.(\ref{38A}). The form
factors are summarized in Eqs.(\ref{42}-\ref{44}) and their values at $%
q^{2}=0$ are given in Tabel-I. Using these form factors we studied the
observables, i.e. the branching ratio and helicity fraction of $D_{s}^{\ast }
$ in the decay $B_{c}\rightarrow D_{s}^{\ast }l^{+}l^{-}$ $(l=e,\mu )$ both
in SM and in ACD\ model, which has one additional parameter i.e. the inverse
compactification radius $1/R.$ The effects of extra dimensions to the
helicity fraction of $D_{s}^{\ast }$ is very mild for the case when the
tauon $(\tau )$ is taken as a final state lepton as shown in fig 3, however
the effects of extra dimensions are quite significant for the case when muon
($\mu $) is taken as a final state lepton as shown in fig 2. In near future
when \textbf{LHC }is fully operational where more data is available, will put
a stringent constraint on compactification radius $R$ and gives us a deep
understanding of $B$ Physics.

\section*{Acknowledgements}

The authors would like to thank Profs. Riazuddin and Fayyazuddin for their
valuable guidance and helpful discussions. The authors M. A. P. and M. J. A.
would like to acknowledge the facilities provided by National Centre for
Physics during this work.

\end{document}